\documentclass[lettersize,journal]{IEEEtran}
\usepackage{subfigure}
\usepackage[table,xcdraw]{xcolor}
\usepackage{ulem}
\usepackage{url}
\usepackage{algorithm}
\usepackage{algpseudocode}
\usepackage{amsmath}
\usepackage{amsfonts}
\usepackage{booktabs}
\usepackage{multirow}
\usepackage{enumitem}

\usepackage{caption}

\usepackage{graphicx}

\usepackage[
colorlinks=true, % Reference color setup	
linkcolor=black, % blue
citecolor=black, % blue
urlcolor=black   % magenta
]{hyperref}

\newcommand{\figname}{Fig.} 

\begin{document}
\title{EncodingNet: A Novel Encoding-based MAC Design for Efficient Neural Network Acceleration}

%\author{IEEE Publication Technology,~\IEEEmembership{Staff,~IEEE,}
        % <-this % stops a space
%\thanks{This paper was produced by the IEEE Publication Technology Group. They are in Piscataway, NJ.}% <-this % stops a space
%\thanks{Manuscript received April 19, 2021; revised August 16, 2021.}
%}
\author{Bo Liu, Bing Li, \IEEEmembership{Senior Member, IEEE}, Grace Li Zhang, \IEEEmembership{Member, IEEE}, Xunzhao Yin, \IEEEmembership{Member, IEEE}, Cheng Zhuo, \IEEEmembership{Senior Member, IEEE}, and Ulf Schlichtmann, \IEEEmembership{Senior Member, IEEE}
%       Institute for Electronic Design Automation, Technische Universitaet Muenchen, Germany\\
%       \{b.li, ning.chen, ulf.schlichtmann\}@tum.de
%       }
\thanks{This work is funded by the Deutsche Forschungsgemeinschaft (DFG, German Research Foundation) – Project-ID 504518248 and by TUM International Graduate School of Science and Engineering (IGSSE).}

\thanks{Bo Liu and Ulf Schlichtmann are 
with the Chair of Electronic Design Automation, Technical University of Munich, 80333 Munich, Germany
(e-mail: bo.liu@tum.de; ulf.schlichtmann@tum.de).}
\thanks{Bing Li is with the Digital Integrated Systems Group, University of Siegen, 57068 Siegen, Germany (e-mail:bing.li@uni-siegen.de).}
\thanks{Grace Li Zhang is with the research group of Hardware for Artificial Intelligence, Technical University of Darmstdat, 64283 Darmstadt, Germany (e-mail: grace.zhang@tu-darmstadt.de).}
\thanks{Xunzhao Yin and Cheng Zhuo are with Zhejiang University, 310027 Hangzhou, China (e-mail: xzyin1@zju.edu.cn; czhuo@zju.edu.cn).}
}
\markboth{IEEE Transactions on Circuits and Systems for Artificial Intelligence}
{Shell \MakeLowercase{\textit{et al.}}: IEEE Transactions on Circuits and Systems for Artificial Intelligence}


\maketitle

\begin{abstract}%
Deep neural networks (DNNs) have achieved great breakthroughs in many fields such as image classification and natural language processing.
However, the execution of DNNs needs to conduct massive numbers of multiply-accumulate (MAC) operations on hardware and thus incurs a large power consumption. To address this challenge, we propose a novel digital MAC design based on encoding. In this new design, the multipliers are replaced by simple logic gates to
% \textcolor{blue}{represent the results with a wide bit representation}.
represent the results with a wide bit representation.
%These bits carry individual position weights 
%\textcolor{blue}{\sout{, which can be trained for specific neural networks to enhance inference accuracy}.} 
%The wide bits and the corresponding position weights are then used to calculate the outputs of neurons by bit-wise weighted accumulation in an MAC array. 
The outputs of the new multipliers are added by bit-wise weighted accumulation and the accumulation results are compatible with existing computing platforms accelerating 
neural networks.
%\textcolor{blue}{\sout{with either uniform or non-uniform quantization}}. 
Since the multiplication function is
% \textcolor{blue}{replaced by a simple logic representation}
replaced by a simple logic representation, the critical paths in the resulting circuits become much shorter. Correspondingly, pipelining stages and intermediate registers used to store partial sums in the MAC array can be reduced, leading to a significantly smaller area as well as better power efficiency.
The proposed design has been synthesized and verified by 
ResNet18-Cifar10, ResNet20-Cifar100, ResNet50-ImageNet, MobileNetV2-Cifar10, MobileNetV2-Cifar100, and EfficientNetB0-ImageNet. 
The experimental results confirmed the reduction of circuit area by up to 48.79\% and the reduction of power consumption of executing DNNs by up to 64.41\%, while the accuracy of the neural networks can still be well maintained. 
%\textcolor{blue}{Code will be available on GitHub after the paper's acceptance.
%}
% \textcolor{blue}{
The open source code of this work can be found on GitHub with link
``\url{https://github.com/Bo-Liu-TUM/EncodingNet/}''.
% }
\end{abstract}


% \begin{table*}[]
% \centering
% \caption{hardware}
% \begin{tabular}{ccccccccc}
% \toprule
% \multirow{2}{*}{\begin{tabular}[c]{@{}c@{}}Size of Systolic Array\end{tabular}} &
%   \multicolumn{2}{c}{Bit-Width of Product} &
%   \multicolumn{3}{c}{Power of Systolic Array (W)} &
%   \multicolumn{3}{c}{Area of Systolic Array ($mm^2$)} \\ 
%  &
%   \multicolumn{1}{c}{Traditional} &
%   Proposed &
%   \multicolumn{1}{c}{Traditional} &
%   \multicolumn{1}{c}{Proposed} &
%   Reduced &
%   \multicolumn{1}{c}{Traditional} &
%   \multicolumn{1}{c}{Proposed} &
%   Reduced \\ \hline
% 64$\times$64   & \multicolumn{1}{c}{16} & 48 & \multicolumn{1}{c}{0.65} & \multicolumn{1}{c}{0.40} & 38.07\% & \multicolumn{1}{c}{0.89} & \multicolumn{1}{c}{0.42} & 53.36\% \\ 
% 128$\times$128 & \multicolumn{1}{c}{16} & 48 & \multicolumn{1}{c}{2.46} & \multicolumn{1}{c}{1.05} & 57.38\% & \multicolumn{1}{c}{3.43} & \multicolumn{1}{c}{1.04} & 69.61\% \\ 
% 256$\times$256 & \multicolumn{1}{c}{16} & 48 & \multicolumn{1}{c}{9.57} & \multicolumn{1}{c}{2.85} & 70.18\% & \multicolumn{1}{c}{13.50} & \multicolumn{1}{c}{2.74} & 79.63\% \\ 
% \bottomrule
% \end{tabular}
% \end{table*}

\section{Introduction}
The last decade has witnessed the success of deep neural networks (DNNs) in many fields, e.g., image classification and speech recognition. DNNs achieved this success by executing huge numbers of multiply-accumulate (MAC) operations. %For example, ResNet152 has 152 layers, resulting in about 60 million weights and 11.3 billion of MAC operations \cite{???}. 
Executing such massive numbers of MAC operations requires a huge amount of dedicated hardware resources and incurs large power consumption.  For example, GPT-3 used in ChatGPT \cite{online} has 96 layers with 175 billion weights for the synapses \cite{language}. This results in trillions of MAC operations to be executed. To use GPT-3, 10,000 V100 Graphics Processing Units (GPUs) were used \cite{carbon}. It was estimated that training GPT-3 consumed 1287 MWh energy \cite{carbon}, which is comparable to the electricity consumption of 120 years for an average U.S. household \cite{online1}.

%and thus preventing their applications in resource-constrained platforms, e.g., edge devices.
%However, a major challenge when deploying DNNs on hardware is high power consumption\cite{???}. 
%The enormous power consumption of DNNs mainly comes from data movement and data computation in %processing MAC operations on digital hardware \cite{???}. 

%two aspects: a huge number of multiply-and-accumulate (MAC) operations, and frequent data movement\cite{???}. 
%In well-known DNN models, more than 99\% DNN operations are MAC operations \cite{sharma2018bit}. 
%Despite the design of excellent hardware architectures such as systolic array \cite{???} to balance the power consumption, computational speed, and chip area of MAC operations, 
%the power consumption remains high.
%The power consumption caused by data movement becomes more severe as the model size becomes larger.
%These problems make it challenging to deploy DNNs on resource-constrained edge devices.

% 为了解决功耗问题，主要有以下几大类方法：
% 1、pruning，
% 2、quantization，定点 approximate 浮点
% 3、mac suspension，clock gating 和 power gating
% 4、DVFS，不同层不同量化bit数来调整电压和频率，低比特数乘法需要较低的电压
% 1是减少乘法的数量，2-4是针对mac电路本身做优化
%To address the enormous power consumption of DNNs executed on hardware, one research direction is
Various techniques have been proposed to enhance the execution efficiency of DNNs on digital hardware. For example, efficient data flows, e.g., weight-stationary \cite{TPU}, output stationary \cite{Diannao}, and row-stationary \cite{eyriss} have been introduced to reduce data movement in executing MAC operations.
% Another direction is to reduce the power consumption incurred by MAC computation from both software perspective and  hardware perspective. 
%From the software perspective, previous work tries to reduce the number of MAC operations in DNNs required to be executed on hardware. For example, 
Pruning has been deployed to compress DNNs by pruning unnecessary weights \cite{pruning,LIANG2021370,10546870,PZCS23}. 
Knowledge distillation \cite{distilling,10546606} transfers a large DNN model to a compact model consisting of few MAC operations. 
In addition, dynamic neural networks \cite{dynamic,10595861} skip MAC operations to make decisions early instead of reaching the output layers of DNNs. Furthermore, neural architecture search (NAS) \cite{Wistuba2019ASO} has been explored extensively to obtain efficient neural network structures %automatically.  
with few MAC operations.
% \textcolor{blue}{\sout{automatically}}

From the hardware perspective, MAC approximation, MAC suspension and MAC voltage/frequency scaling have been applied to enhance the computational and power efficiency of executing MAC operations. Specifically, approximate computing \cite{approximate} allows inaccuracy in MAC operations so that the logic complexity and thus power consumption can be reduced. Quantization \cite{quantization,10137171} approximates floating-point MAC operations with fixed-point operations to reduce logic complexity and thus power consumption.  %These methods inevitably cause an accuracy drop of DNNs, which can be compensated by retraining. While MAC approximation changes the logic circuits of the MAC units, 
MAC suspension disables MAC units in a hardware accelerator when they are not used. For example, the multipliers with the weights equal to 0 can be disabled by clock/power gating \cite{powergating,clockgating}. %The former disables flip-flops in the MAC units to avoid unnecessary signal switches and the latter disconnects supply voltages to disable whole MAC units, so that power consumption can be reduced on-the-fly according to the weights loaded into the MAC array. Different from MAC suspension, 
MAC voltage/frequency scaling \cite{voltagescaling,8531784,PZCS23} adjusts voltage/frequency of MAC units dynamically according to, e.g., required quantization bits and selected weights, to reduce energy consumption while maintaining the functionality of MAC units and inference accuracy.

Despite the active research on efficient neural networks and their execution on hardware, most state-of-the-art work is still restricted by the assumption that multipliers and adders for executing MAC operations are designed based on traditional logic design, where
%encoding technique where 
the results of multiplication and addition are represented with the two’s complement binary number system \cite{von1993first,lyon1976two}. 
%
%This assumption unnecessarily confines the design space of MAC units %and can thus lead to a large area and power consumption.
For example, EvoApprox8b \cite{mrazek2017evoapprox8b}, an open-source library, contains a variety of circuits of unsigned 8-bit approximate multipliers and adders that have been optimized by Multi-objective Cartesian Genetic Programming \cite{hrbacek2016automatic} to achieve Pareto optimal trade-offs between power, area and approximate error. 
Approximate multipliers in EvoApprox8b have been utilized in neural networks \cite{pinos2023acceleration, mrazek2020libraries}, achieving acceptable accuracy while also reducing area and power consumption.
However, since these
% the outputs of these approximate multipliers and adders 
designs 
%in the EvoApprox8b 
still rely on two's complement format, their design space is limited, resulting in only modest reductions in area and power consumption of MAC arrays.
%The execution of MAC operations on such multipliers and adders is not power-efficient. 
%Accordingly, the hardware acceleration of MAC operations is still restricted by the binary number systems and encoding format used to represent data. 
Although some previous work has attempted to take advantage of new data encoding for efficiently executing MAC operations, e.g., logarithmic number system \cite{miyashita2016convolutional} and residue number system \cite{RNS2020}, they still suffer from a large cost due to data conversion into binary number system.

\begin{figure}[t]
\centering
\includegraphics{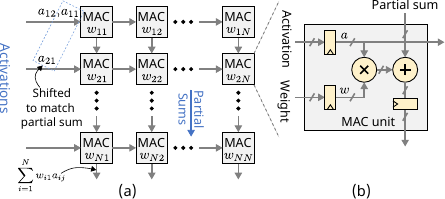}
  \caption{(a) Structure of systolic array according to
  \cite{TPU}. (b)
  Structure of an MAC unit.}
\label{fig:SystolicArray}
\end{figure}

Different from previous work, we propose a novel digital MAC design by directly exploring the encoding to simplify MAC circuits for efficient DNN acceleration.  
%with which hardware platforms can be constructed to efficiently execute DNNs.
%. Such MAC designs are then used to construct a hardware platform to executed MAC operations more power-efficiently. 
%Different from the various number systems mentioned above, 
%most of them either only focus on binary number system, 
%or propose a number system with challengs of data conversion, 
%we propose to use generalized weighted number system (GWNS, explained in Section ???) to represent the product of weights and activations and their partial sum, 
%because of its low complexity of data conversion and custom encoding format property. 
The key contributions are summarized as follows:

\begin{itemize}
%[topsep=8pt,itemsep=2pt,leftmargin=10pt]

\item The encoding at the outputs of multipliers in MAC units is 
% \textcolor{blue}{
% \sout{examined directly to simplify}
% determined in such a way that
% }
determined in 
such 
a way that
the logic of the multipliers
% \textcolor{blue}{can be simplified}
can be simplified. 
With different encodings, the resulting logic can deviate from the traditional logic function of multipliers but lead to significantly simpler circuit implementation. This perspective opens up %the new potential for future techniques to 
a new dimension to 
search for more efficient logic of MAC operations 
% \textcolor{blue}{to accelerate DNNs efficiently}
to accelerate DNNs efficiently
beyond existing arithmetic expressions of multiplication and addition functions.
% \textcolor{blue}{\sout{to accelerate DNNs efficiently}}

%A novel digital MAC design based on a completely new encoding is proposed. In this new design, the multipliers are replaced by simple logic gates to project the results to a wide bit representation. This simple logic projection bring shorter critical paths in the resulting circuits.

\item The simple logic 
% implementing the 
mapping from the inputs of the multipliers to their outputs is determined by Cartesian Genetic Programming (CGP) \cite{miller1997designing,miller2015cartesian} to approximate the original output values. The bit width of the encoded outputs is much wider than that of the original multiplier outputs, leading to a low-complexity logic mapping. 
% Therefore, the logic complexity of this mapping becomes much low due to this projection of outputs onto wide bits. 
%The exact width of the encoding of the multipliers is determined by a binary search to balance inference accuracy and power consumption as well as area cost. 

%The encoding of the redesigned multiplier is determined by a binary search %algorithm while balancing  inference accuracy and power consumption as well as %area cost. Since
%the wide bits of the multiplication results  in the new encoding carry individual %position weights, they are trained for specific neural networks to further %enhance inference accuracy. 

\item The wide bits at the outputs of the encoding-based multipliers carry individual position weights, which are determined by the logic circuit implementing the multiplier and the encoding.
%\textcolor{blue}{\sout{, which are trained for specific neural networks to enhance inference accuracy}}. 
The wide bits and the corresponding position weights are used to calculate the outputs of 
MAC operations of a neuron
by bit-wise weighted accumulation in a MAC array. 
These outputs 
are in the original 
two's complement formats, 
%with either uniform or non-uniform quantization, 
so that the proposed design is compatible with 
peripheral circuits in existing computing systems.

\item %The redesigned MAC design can be used to construct general-purpose hardware platforms or task-specific platforms.
% \textcolor{blue}{\sout{Since the critical paths in the encoding-based MAC design become much shorter, pipelining stages in the MAC array with these simplified circuits can be reduced significantly, which can be taken advantage of to reduce the area and power consumption of the MAC array. }}
% \textcolor{blue}{
 The encoding-based multipliers, characterized by fewer logic levels, in conjunction with highly parallelized bit-wise accumulators, considerably shorten the critical paths in the encoding-based MAC array design. Therefore, pipelining stages and intermediate registers used to store partial sums can be significantly reduced, leading to a much smaller area and lower power consumption.
\end{itemize}

The rest of the paper is structured as follows. Section \ref{sec:Motivation} explains the motivation of this work. %Section \ref{sec:concept} describes the concept of the encoding-based MAC design. 
Section \ref{sec:concept} 
elaborates the details of the proposed encoding-based MAC design. 
Experimental results are presented in Section \ref{sec:results} and conclusions are drawn in Section \ref{sec:conclusion}. 
% \textcolor{blue}{}
All symbols used in the paper and their definitions are listed in Table \ref{tab:notations}.

\section{Concept and Comparison with Related Work}\label{sec:Motivation} 
 In DNNs, there are massive amounts of MAC operations. 
Existing digital hardware platforms use many parallel MAC units, e.g., 65,536 in the systolic array of TPU v1 \cite{TPU}, to accelerate DNNs. 
The structure of this systolic array is sketched in \figname~\ref{fig:SystolicArray}(a), while the internal structure
of a MAC unit is shown in \figname~\ref{fig:SystolicArray}(b).  
In the systolic array, weights are preloaded and
activations are streamed as inputs. The partial sum of a multiplication is propagated along a column to calculate the multiplication result of an input vector and a weight vector. Between rows and columns there are flip-flops. Therefore, the activations are  shifted to match the propagation of the partial sums at the MAC units. 
% The internal structure of an MAC unit is shown in \figname~\ref{fig:SystolicArray}(b).
 
%Inside a MAC unit as shown in \figname~x(b), the multiplier multiplies the input and the weigtht to generate the partial sum. The partial sums are added by the adders to calculate the result at a neuron. 

\begin{figure}
\centering
\subfigure[]{
% \subfloat{
\begin{minipage}[c]{0.5\linewidth}
\setlength{\tabcolsep}{1.5pt}%2.5pt
\renewcommand{\arraystretch}{1.0}
\centering
\scriptsize

\begin{tabular}{ccccc}
\toprule
% \multirow{2}{*}{In1} & \multirow{2}{*}{In2} & \multicolumn{2}{c}{Out Encoding} &    \\
In1 
% \multirow{2}{*}{In1}
& 
In2 
% \multirow{2}{*}{In2}
& Trad. Enc.          & New Enc.           & Value \\
\text{\tiny$I_1^1I_0^1$}

& 
\text{\tiny$I_1^2I_0^2$} 

& \text{\tiny$b_3b_2b_1b_0$}  & \text{\tiny$b_4b_3b_2b_1b_0$}  & $v$     \\\midrule
10                   & 10                   & 0100            & 01111          & 4     \\
10                   & 11                   & 0010            & 00111          & 2     \\
10                   & 00                   & 0000            & 11111          & 0     \\
10                   & 01                   & 1110            & 10111          & -2    \\
11                   & 10                   & 0010            & 01011          & 2     \\
11                   & 11                   & 0001            & 00001          & 1     \\
11                   & 00                   & 0000            & 11111          & 0     \\
11                   & 01                   & 1111            & 10101          & -1    \\
00                   & 10                   & 0000            & 11111          & 0     \\
00                   & 11                   & 0000            & 11111          & 0     \\
00                   & 00                   & 0000            & 11111          & 0     \\
00                   & 01                   & 0000            & 11111          & 0     \\
01                   & 10                   & 1110            & 11011          & -2    \\
01                   & 11                   & 1111            & 11001          & -1    \\
01                   & 00                   & 0000            & 11111          & 0     \\
01                   & 01                   & 0001            & 11101          & 1     \\\bottomrule
\multicolumn{5}{l}{\tiny
\begin{tabular}[c]{@{}l@{}}
\vspace{-0.15cm}
\\
$v=\sum_{i=0}^{M-1} s_{i}  b_{i}$, where $s_{i}$ is position weight. \\
Trad.:  $M=4, s_{3}=-8, s_{2}=4, s_{1}=2,$ \\
\ \ \ \ \ \ \ \ $s_{0}=1$  \\
New:  $M=5, s_{4}=-4, s_{3}=2, s_{2}=2,$ \\ 
\ \ \ \ \ \ \ \ $s_{1}=-1, s_{0}=1  $
\end{tabular}}                 
\end{tabular}
\label{tab:mul-tt}
\end{minipage}%
}%
\hspace{0.0001\linewidth}
\begin{minipage}[c]{0.45\linewidth}
\centering
\subfigure[]{
% \subfloat[]{
\begin{minipage}[c]{1\linewidth}
\centering
\label{fig:mul-trad}
\includegraphics[width=1\linewidth]{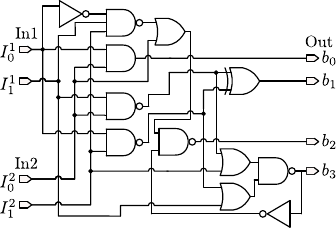}
\end{minipage}%
}%
\vspace{0.1\linewidth}
\subfigure[]{
% \subfloat[]{
\begin{minipage}[c]{0.6\linewidth}
\centering
\label{fig:mul-our}
\includegraphics[width=1\linewidth]{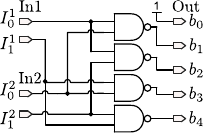}
\end{minipage}%
}%
\end{minipage}%
\caption{
% \textcolor{blue}{Truth tables and corresponding multipliers with traditional two's complement encoding and a new encoding.}
Truth tables and corresponding multipliers with traditional two's complement encoding and a new encoding.
(a) Truth tables of multipliers with the traditional encoding and a new encoding, 
% \textcolor{blue}{where the position weights are shown at the bottom}
where the position weights are shown at the bottom.
(b) The traditional 2-bit signed multiplier. (c) The multiplier with a new encoding, 
% \textcolor{blue}{called an \textit{encoding-based multiplier}}
called an \textit{encoding-based multiplier}. 
%acomparison between traditional design in binary system (two's complement) and an example of proposed design in our encoding. (a) MAC in binary. (b) MAC in our encoding.
%Illustration of proposed binary search algorithm for the encoding of the product of 8-bit uniform quantization levels. Left: RMSE vs. iterations of search algorithm. Right: RMSE vs. the number of random samples.
}
\label{fig:MulComp}
\end{figure}

In a MAC unit above, the inputs of the multiplier are represented in the two's complement to express integer values. The circuit of the multiplier is defined by the truth table which enumerates all the input combinations.    
%These MAC units rely on traditional digital multipliers and adders. % which are 
% NN中有很多乘加运算，用MAC单元来加速乘加运算，MAC单元构成一个阵列，并行地执行乘加运算
%The massive amount of multiply-and-accumulate (MAC) operations is a major challenge when executing DNNs on hardware. To address this issue, hardware architectures consisting of MAC units such as systolic array have been designed for acceleration.
% 介绍传统MAC中的乘法器是怎么设计的，基于二进制补码，用图中2bit有符号数乘法器为例子
%Traditional MAC units consist of multipliers and adders. 
%These multipliers and adders are designed based on the traditional encoding technique where 
% the results of multiplication and addition are evaluated and represented within the %confines of the two's-complement number system. 
For example, \figname~\ref{fig:MulComp}(a) shows the truth table of a multiplier with 2-bit signed inputs \textit{In1} and \textit{In2}. The column \textit{Trad. Enc.} shows the output bit sequences in the two's complement format
corresponding to decimal numbers in the last column of \figname~\ref{fig:MulComp}(a). From this truth table, the logic circuit for this multiplier can be synthesized as shown in \figname~\ref{fig:MulComp}(b). %\textcolor{red}{This circuit is more complex than that in https://en.wikipedia.org/wiki/Binary\_multiplier.} 
As the bit-width of input operands increases, the number of rows in the truth table of a multiplier increases exponentially. Since the synthesized circuit must be able to realize all the rows in the truth table exactly, the circuit thus becomes complicated quickly. For example, an 8-bit signed multiplier can contain 417 combinational logic gates. Though approximate computing \cite{approximate, mrazek2017evoapprox8b, pinos2023acceleration,mrazek2020libraries} can be applied to reduce the logic complexity of multipliers, this technique still uses the two's complement format to represent the multiplication results %is still based on the original logic function of multiplier 
and does not take advantage of the full potential of MAC units.

The circuit of a multiplier maps the input combinations to the output combinations. 
%Accordingly, the bit sequence used at the output of the multiplier to represent integer numbers can affect the circuit complexity. 
In the traditional design, the bit sequences representing the output combinations of a multiplier are 
predefined in the two's complement format 
according to the multiplication function, as shown
%For example, the bit sequences 
in the \textit{Trad. Enc.} column in \figname~\ref{fig:MulComp}(a). 
%correspond to the signed representations of multiplication results of 2-bit signed inputs. 
However, if these bit sequences can be adjusted, the new truth table can lead to a multiplier circuit with a lower logic complexity. 
For example, the \textit{New Enc.} column in \figname~\ref{fig:MulComp}(a) shows another assignment of bit sequences to represent the same output values of the multiplier, where the bit width has been increased from 4 to 5.
Since the number of bits at the output of the multiplier has been increased, different bit sequences can represent a same integer value. For example, both
00111 and 01011 in the \textit{New Enc.} column in \figname~\ref{fig:MulComp}(a) represent
% the same
decimal value 2.
%In addition, the number of bits at the output of multiplier can also be expanded, which can provide more flexibitily to choose the bit sequence at the output of the multiplier for logic simplification. 
From this new bit sequence assignment, a much simpler circuit can be generated, as illustrated in \figname~\ref{fig:MulComp}(c).
% \textcolor{blue}{
Please note that both \textit{Trad. Enc.} and \textit{New Enc.} have position weights, as shown at the bottom of \figname~\ref{fig:MulComp}(a).

\begin{table}[t]
\setlength{\tabcolsep}{2.0pt}%2.5pt
\renewcommand{\arraystretch}{1.0}
\centering
\footnotesize
\caption{List of symbols and their definitions.}
\begin{tabular}{ccc}
\toprule
Symbol                                 & Range            & Definition                                                 \\ \midrule
$B$                                    & set              & $B=\{0, 1\}$                                               \\
$b_3b_2b_1b_0$                         & vector           & 4-bit bit sequence, each $b \in B$                                         \\
$r$                                    & $\mathbb{N}^+$   & logic depths (the number of rows of internal nodes)        \\
$c$                                    & $\mathbb{N}^+$   & logic levels (the number of columns of internal nodes)     \\
$n$                                    & $\mathbb{N}^+$   & the number of input nodes                                  \\
$m$                                    & $\mathbb{N}^+$   & the number of output nodes                                 \\
$M$                                    & $\mathbb{N}^+$   & output bit-width of a candidate multiplier, $M \le m$      \\
$N$                                    & $\mathbb{N}^+$   & size of systolic/MAC array ($N \times N$)                  \\
$\textbf{\textit{s}}, \textbf{\textit{s}}^{*}$                  & vector           & position weights                                           \\
$\textbf{\textit{v}}$                  & vector           & outputs of exact multiplier w.r.t to all inputs            \\
$\textbf{\textit{B}}$                  & matrix           & candidate multiplier output bits w.r.t to all inputs       \\
$\varepsilon$                          & $\mathbb{Q}^+$   & maximal relative error                                     \\
$\varepsilon_{th}$                     & $\mathbb{Q}^+$   & maximal relative error threshold                           \\
$\Gamma$                               & set              & gate library                                               \\
$\mathcal{C}$                          & function         & cost function                                           \\
$\mathcal{A}_{max}$                    & function         & maximum area of all possible candidate multipliers         \\
$\mathcal{A}_{\mathcal{M}^{\prime}}$   & function         & area of a candidate multiplier                             \\
$x, y, z, z^{\prime}$                  & vector           & Boolean vector with given bits                             \\
$\delta$, $\delta^{\prime}$            & function         & assign a natural number to a Boolean vector                \\
$\mathcal{M}$, $\mathcal{M}^{\prime}$  & function         & map a Boolean  vector to another Boolean vector            \\
\bottomrule
\end{tabular}
\label{tab:notations}
\end{table}

The bit sequence assignment in \figname~\ref{fig:MulComp}(a) is called an \textit{encoding}.
The original encoding of the multiplier shown in the column \textit{Trad. Enc.} is only one of the possible encodings representing the values at the output of the multiplier.
Since
% \textcolor{blue}{
% \sout{various}
different
% }
encodings lead to different truth tables for the multiplier, they also result in different circuit complexity after logic synthesis.
% \textcolor{blue}{
The multiplier with a new encoding rather than two's complement at the output is called an \textit{encoding-based multiplier}.
Accordingly, the MAC array constructed by encoding-based multipliers and adders is called an \textit{encoding-based MAC array}.
% }
Therefore, exploring the encoding can be an effective technique to obtain more efficient circuit implementations for the multiplier and MAC array.

%\subsection{Comparison of Related Work}

% \textcolor{blue}{
Previous work related to the proposed concept majorly focuses on exploring new data representations, optimization of multipliers and MAC units, and approximate computing applied to DNNs to reduce area and power consumption. 
% }

\begin{table}
{
\setlength{\tabcolsep}{1.0pt}%2.5pt
\renewcommand{\arraystretch}{1.0}
\centering
% \tiny
\fontsize{6.5}{6.5} \selectfont
\caption{Comparison with state-of-the-art designs}
\begin{tabular}{cccccccc}
\toprule
  \multicolumn{1}{c}{\begin{tabular}[c]{@{}c@{}}Work\end{tabular}}
& \multicolumn{1}{c}{\begin{tabular}[c]{@{}c@{}}Multiplier \\Type\end{tabular}}
& \multicolumn{1}{c}{\begin{tabular}[c]{@{}c@{}}Multiplier \\Output \\Format\end{tabular}}
& \multicolumn{1}{c}{\begin{tabular}[c]{@{}c@{}}Multiplier \\Output \\Bit-Width\end{tabular}}
& \multicolumn{1}{c}{\begin{tabular}[c]{@{}c@{}}Dataset\end{tabular}}
& \multicolumn{1}{c}{\begin{tabular}[c]{@{}c@{}}Model\end{tabular}}
& \multicolumn{1}{c}{\begin{tabular}[c]{@{}c@{}}Acc\textsubscript{A} (\%)\textsuperscript{a}\end{tabular}}
& \multicolumn{1}{c}{\begin{tabular}[c]{@{}c@{}}Acc\textsubscript{loss} (\%)\textsuperscript{c}\end{tabular}}
\\ \midrule
  \multicolumn{1}{c}{\begin{tabular}[c]{@{}c@{}}DAC'21\\\cite{Fibonacci}\end{tabular}}
& \multicolumn{1}{c}{\begin{tabular}[c]{@{}c@{}}Approx \\8b$\times$8b\end{tabular}}
& \multicolumn{1}{c}{\begin{tabular}[c]{@{}c@{}}2's comp.\end{tabular}}
& \multicolumn{1}{c}{\begin{tabular}[c]{@{}c@{}}16b\end{tabular}}
& \multicolumn{1}{c}{\begin{tabular}[c]{@{}c@{}}Cifar100\end{tabular}}
& \multicolumn{1}{c}{\begin{tabular}[c]{@{}c@{}}ResNet18\end{tabular}}
& \multicolumn{1}{c}{\begin{tabular}[c]{@{}c@{}}73.54\end{tabular}}
& \multicolumn{1}{c}{\begin{tabular}[c]{@{}c@{}}-1.74\end{tabular}}
\\ \midrule
  \multicolumn{1}{c}{\begin{tabular}[c]{@{}c@{}}DATE'22\\\cite{G-Thermal-2022}\end{tabular}}
& \multicolumn{1}{c}{\begin{tabular}[c]{@{}c@{}}Approx \\8b$\times$8b\end{tabular}}
& \multicolumn{1}{c}{\begin{tabular}[c]{@{}c@{}}2's comp.\end{tabular}}
& \multicolumn{1}{c}{\begin{tabular}[c]{@{}c@{}}16b\end{tabular}}
& \multicolumn{1}{c}{\begin{tabular}[c]{@{}c@{}}ImageNet\end{tabular}}
& \multicolumn{1}{c}{\begin{tabular}[c]{@{}c@{}}ResNet50\end{tabular}}
& \multicolumn{1}{c}{\begin{tabular}[c]{@{}c@{}}72.5\end{tabular}}
& \multicolumn{1}{c}{\begin{tabular}[c]{@{}c@{}}-\end{tabular}}
\\ \midrule
  \multicolumn{1}{c}{\begin{tabular}[c]{@{}c@{}}TC'22\\\cite{P-Learning-2022}\end{tabular}}
& \multicolumn{1}{c}{\begin{tabular}[c]{@{}c@{}}Approx \\8b$\times$8b\end{tabular}}
& \multicolumn{1}{c}{\begin{tabular}[c]{@{}c@{}}2's comp.\end{tabular}}
& \multicolumn{1}{c}{\begin{tabular}[c]{@{}c@{}}16b\end{tabular}}
& \multicolumn{1}{c}{\begin{tabular}[c]{@{}c@{}}Cifar100 \\ImageNet\end{tabular}}
& \multicolumn{1}{c}{\begin{tabular}[c]{@{}c@{}}MobileNetV2 \\ResNet18\end{tabular}}
& \multicolumn{1}{c}{\begin{tabular}[c]{@{}c@{}}-\\-\end{tabular}}
& \multicolumn{1}{c}{\begin{tabular}[c]{@{}c@{}}-2.4\\-1.3\end{tabular}}
\\ \midrule
  \multicolumn{1}{c}{\begin{tabular}[c]{@{}c@{}}VLSI'24\\\cite{T-Toward-2024}\end{tabular}}
& \multicolumn{1}{c}{\begin{tabular}[c]{@{}c@{}}Approx \\8b$\times$6b\end{tabular}}
& \multicolumn{1}{c}{\begin{tabular}[c]{@{}c@{}}2's comp.\end{tabular}}
& \multicolumn{1}{c}{\begin{tabular}[c]{@{}c@{}}16b\textsuperscript{b}\end{tabular}}
& \multicolumn{1}{c}{\begin{tabular}[c]{@{}c@{}}Cifar100\\Cifar100\\ImageNet\\ImageNet\end{tabular}}
& \multicolumn{1}{c}{\begin{tabular}[c]{@{}c@{}}ResNet18\\MobileNetV2\\ResNet18\\ResNet50\end{tabular}}
& \multicolumn{1}{c}{\begin{tabular}[c]{@{}c@{}}76.11\\67.76\\68.28\\75.0 \end{tabular}}
& \multicolumn{1}{c}{\begin{tabular}[c]{@{}c@{}}-0.27\\-1.36\\-1.48\\-1.13\end{tabular}}
\\ \midrule
  \multicolumn{1}{c}{\begin{tabular}[c]{@{}c@{}}ours\end{tabular}}
& \multicolumn{1}{c}{\begin{tabular}[c]{@{}c@{}}Approx \\8b$\times$8b\textsuperscript{b}\end{tabular}}
& \multicolumn{1}{c}{\begin{tabular}[c]{@{}c@{}}searched \\encoding\end{tabular}}
& \multicolumn{1}{c}{\begin{tabular}[c]{@{}c@{}}64b\textsuperscript{b}\end{tabular}}
& \multicolumn{1}{c}{\begin{tabular}[c]{@{}c@{}}
Cifar-10\\
Cifar-100\\
ImageNet\\
Cifar-10\\
Cifar-100\\
ImageNet
\end{tabular}}
& \multicolumn{1}{c}{\begin{tabular}[c]{@{}c@{}}
ResNet18\\
ResNet20\\
ResNet50\\
MobileNetV2\\
MobileNetV2\\
EfficientNet-B0
\end{tabular}}
& \multicolumn{1}{c}{\begin{tabular}[c]{@{}c@{}}
93.14\\
68.32\\
75.46\\
93.74\\
71.23\\
69.96
\end{tabular}}
& \multicolumn{1}{c}{\begin{tabular}[c]{@{}c@{}}
+0.07\\
-0.5\\
-1.59\\
-0.17\\
+0.06\\
-7.73
\end{tabular}}
\\
\bottomrule

\multicolumn{8}{l}{\begin{tabular}[l]{@{}l@{}}
\vspace{-0.15cm}
\\
\textsuperscript{a} Top-1 accuracy when using approximate multiplier \\
\textsuperscript{b} 8b means 8-bit, 6b means 6-bit, 16b means 16-bit, and 64b means 64-bit \\
\textsuperscript{c} Top-1 accuracy loss compared with the 32-bit floating-point model
\end{tabular}}\\

\end{tabular}
\label{tab:sota-designs-compare-2}

}
\end{table}

% \textcolor{blue}{
On novel data representations, in \cite{miyashita2016convolutional}, weights and activations are represented with a power of 2 by using binary logarithmic quantization. Multiplication operations are replaced by addition of exponents and bit-shifting, eliminating the need for digital multipliers and enhancing computational efficiency. However, the logarithmic number system is less efficient to represent the distribution of weights and activations, which confines the accuracy in lower bit-width data representation.
Residue number system (RNS) has also been explored to optimize the hardware implementation of convolutional layers \cite{RNS2020}. Multiplications and accumulations in traditional 2’s complement are replaced with the counterparts in RNS without carry propagation, which speeds up operations and improves the computational efficiency. But RNS suffers from a large cost of modular operations and data conversion into binary number system. 
Fibonacci encoding technique has been used to encode weights of neural networks \cite{Fibonacci}, while the activations and products are still in 2's complement. 
By using the Fibonacci encoding technique, some full adders in the multipliers can be replaced with OR gates, leading to a reduced area, power and delay. 
However, the Fibonacci encoding technique requires special handling of weights, and restricts the range of representable values (especially, it cannot encode negative values), potentially affecting model expressiveness.
% }

% \textcolor{blue}{
On the optimization of exact multipliers, 
multi-bit recoding is used to optimize multiplication by grouping multiple bits of the multiplier and re-encoding them to reduce the number of partial products, such as the Booth multipier and its variants \cite{Booth}.
To compress the partial product towards less bits representation or less compression layers, schemes such as Wallace tree \cite{Wallace} and Dadda tree \cite{Dadda} are proposed. 
Overlapped multiple-bit scanning multiplication is an optimized multiplication method where multiple bits of the multiplier are processed with overlapping groups to reduce operations and improve speed \cite{OMBSSM}. 
These design schemes are widely applied, but all of them rely on the 2's complement format and do not change the encoding of the multiplication output. 
Further optimization of MAC units include \cite{IMAC-2024} and \cite{Ponraj-2023}, where adders are integrated into multipliers.
% }

% \textcolor{blue}{
On approximate computing, approximate multipliers and adders have been explored in DNN accelerators to reduce the power consumption while maintaining the accuracy \cite{G-Thermal-2022}.  
AutoApprox identifies which layers in a neural network are error-tolerant and implements them with approximate MAC units, while sensitive layers are implemented  with exact MAC units to maintain overall accuracy \cite{P-Learning-2022}. 
ApproxNNs use approximate multipliers to reduce hardware area and energy consumption, and retrain large-scale neural networks to maintain the accuracy %because of the propagation of errors across layers 
with a controlled error \cite{T-Toward-2024}. 
Table \ref{tab:sota-designs-compare-2} compares the proposed approach with these state-of-the-art designs, focusing on widely used datasets and networks.
% }

\section{Encoding-based MAC Design}
\label{sec:concept}

To identify a new encoding to simplify the MAC circuits, two challenges should be addressed. First, the number of encodings is huge, up to $2^{M+16}$ for an 8-bit multiplier with $M$-bit output. For each encoding, a circuit should be generated, which leads to a very long search time. Second, the identified encoding should not make the accumulation implementation 
of partial sums generated by redesigned multipliers complicated. For example, the new encoding shown in the \textit{New Enc.} column in \figname~\ref{fig:MulComp}(a) also defines the input bit combinations of an adder in a MAC operation. However, it is not an easy task to synthesize an efficient circuit for an adder with an arbitrary input encoding.

% \textcolor{red}{
Without a straightforward 
%linear 
relationship between the encoding and the output values of the multiplier, it 
%is necessary 
may need 
to enumerate all possible inputs and outputs to design the multiplier. To reduce design complexity, we do not search the encoding and the circuit of the multiplier separately. Instead, we search for circuit candidates that have small sizes and can maintain the accuracy of the neural networks. These circuit candidates establish the relation betwen the inputs and outputs of the multiplier. Accordingly, the encoding is determined as the results of the circuit candidates
instead of being explicitly specified.

\subsection{Multiplier Design by Structural Search}\label{sec:Methodology-MUL}

\begin{figure}%[htbp]
% \vspace{8pt}
\centerline{\includegraphics[width=0.8\linewidth]{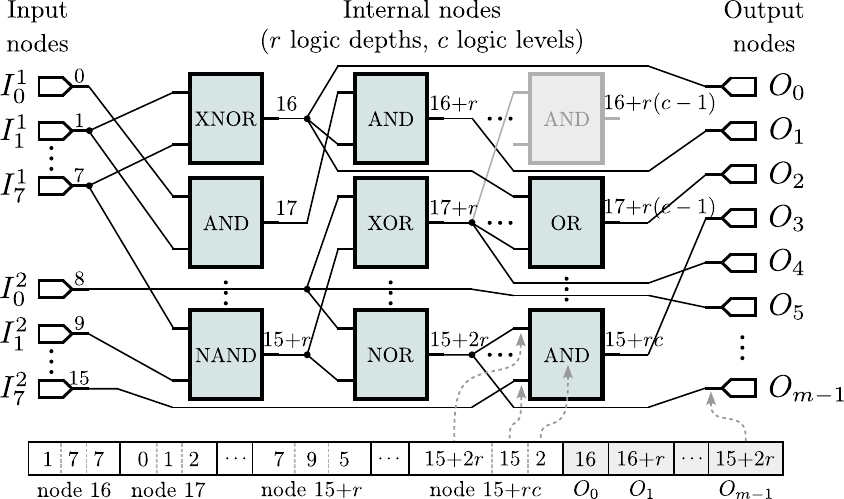}}
\caption{
An example of 8-bit candidate multiplier represented with a 2D regular node array, which consists of $16$ input nodes, $m$ output nodes, and internal nodes with $r$
% \textcolor{blue}{rows (}logic depths\textcolor{blue}{)}
rows (logic depths)
and $c$
% \textcolor{blue}{columns (}logic levels\textcolor{blue}{)}
columns (logic levels).
% \textcolor{blue}{
All input nodes and internal nodes are numbered, i.e., 0, 1, $\dots$, 15, 16, $\dots$, 15+$r$, $\dots$, 15+$rc$. 
The function of each internal node can be any logic in a gate library $\Gamma=$ \{$0^{identity}$, $1^{not}$, $2^{and}$, $3^{or}$, $4^{xor}$, $5^{nand}$, $6^{nor}$, $7^{xnor}$, $8^{const0}$, $9^{const1}$\}, where the numbers, i.e., $0, 1, \dots, 9$, denote the types of the logic gates.
The inputs of internal nodes and output nodes can be connected to previous nodes.
The representation of 2D node array can be coded with a chain of integers, as shown at the bottom.
% }
}
\label{fig_CGP}
\end{figure}

% \textcolor{blue}{
When the input bit-width of a multiplier is small (e.g., 2-bit), the design space is limited, making it possible to find an encoding that accurately represents the multiplier's output using simple logic through exhaustive search.
However, as the bit-width increases (e.g., 8-bit), the search space explodes, making exhaustive search impractical and finding an encoding that can represent the multiplier's output without errors becomes extremely challenging.
% }

% \textcolor{blue}{
To address the above issue, we relax the constraint that the encoding must exactly represent the multiplier's output and 
%introduce 
allow
approximation errors.
Since an encoding system assigns a corresponding output bit sequence to each possible input bit sequence, we can view an encoding as a circuit that maps inputs to outputs.
Then we can use a 2D node array to represent the circuit, as shown in \figname~\ref{fig_CGP}, which consists of input nodes, output nodes and internal nodes.
In this 2D node array representation, each internal node can be one of logic gates, and the inputs of internal nodes and output nodes are connected to other previous nodes, forming a directed acyclic graph.
The process of searching for an encoding is equivalent to continuously altering the logic gates of internal nodes and the connections between nodes till a satisfactory circuit is found.
% }

%To reduce the search space of encoding, 
% \textcolor{red}{
There are two types of encoding to design the multiplier: weighted encoding and non-weighted encoding.
To allow a simple implementation of accumulation, we impose an additional constraint that the bits in a bit sequence have position weights. For example, we assign position weights $s_0$, $s_1$, $s_2$, $s_3$, $s_4$ to the bit sequences in the \textit{New Enc.} column of \figname~\ref{fig:MulComp}(a). Accordingly, a bit sequence  $b_4b_3b_2b_1b_0$ represents the number $\sum_{i=0}^4s_i\times b_i$. These position weights are determined by approximating the function of the original multiplier as accurately as possible.  
%\textcolor{blue}{\sout{These position weights are adjustable for different neural networks to maintain inference accuracy.}} 
Compared with the traditional two's complement number system where the position weights are fixed only to the power of two, %\textcolor{blue}{\sout{the adjustable position weights provide more} the diverse range of values for position weights enhances the} 
the different values of position weights enhance 
flexibility for the implementation of the multipliers and adders. In this paper, we will use weighted encoding since it provides more flexibility in the multiplier design.
% }

% \textcolor{red}{
% \sout{
% Therefore, to search an encoding for the output of an 8-bit signed multiplier, we first view the encoding as a circuit and represent the circuit with a 2D node array.
% Then we apply an algorithm to continuously adjust the logic function of internal nodes and the connections towards a given objective.}
%The following steps are considered:
%1) introduce the 2D node array;
%2) describe the 2D node array;
%% 3) determine the hyperparameters;
%3) apply a search algorithm. 
The 2D node array and the search algorithm to find the circuit for the multiplier are explained in detail below.
% }

\subsubsection{Concept of the 2D Node Array}
As shown in \figname~\ref{fig_CGP}, the 2D node array consists of 16 input nodes, $m$ output nodes and internal nodes with $r$
% \textcolor{blue}{rows (logic depths)}
rows (logic depths)
and $c$
% \textcolor{blue}{columns (logic levels)}
columns (logic levels).
The input number is 16 to match that of the 8-bit original multiplier.
% \textcolor{red}{\sout{
% Generally, all the $m$ output nodes are used as the output bits of the searched encoding-based multiplier, namely $m$ is equal to the number of output bits $M$.
% While in our case, }}
We deliberately set $m$ to be larger than the output bit-width $M$ to expand the exploration space, as the larger number of output nodes can provide more possible exploration of encodings.
% \textcolor{blue}{
Each output node is associated with a position weight, and the absolute value of the position weight represents the importance of that node.
% }
The importance of all output nodes will be evaluated, and only the most $M$ significant nodes will be selected from $m$ as the final output bits of a candidate multiplier.
The function of each internal node can be defined as any logic gate in a logic gate library $\Gamma$.
Each internal node's inputs can be connected to the output of any node in the previous logic levels.

\subsubsection{Description of the 2D Node Array}
% \textcolor{blue}{
As shown in \figname~\ref{fig_CGP}, all the input nodes and internal nodes are numbered, i.e., 0, 1, $\dots$, 15, 16, 17, $\dots$, 15+$r$, 16+$r$, $\dots$, 15+$rc$. These numbers are regarded as addresses that the inputs of internal nodes and output nodes can connect to.
% }
We define a gate library $\Gamma =$ \{$0^{identity}$, $1^{not}$, $2^{and}$, $3^{or}$, $4^{xor}$, $5^{nand}$, $6^{nor}$, $7^{xnor}$, $8^{const0}$, $9^{const1}$\}, where $identity$ represents direct connection without any gate, $const0$ and $const1$ are connected to ground and voltage supply, respectively.
% \textcolor{red}{
The numbers in $\Gamma$, i.e., 0, 1,2, $\dots$, 9, denote the 
%indices 
types
of the logic gates.
% }
Therefore,
% \textcolor{blue}{
the connection and logic function of
% }
each internal node can be coded with three integers, the first two representing which input bits this node connects with and the last integer representing the logic gate type of this node.
For example,
% \textcolor{blue}{
the connection and function of
% }
the internal node 16 can be represented with $[1, 7, 7]$ since its inputs are connected to the 1st and the 7th input node and its logic function is XNOR, whose index in $\Gamma$ is 7.
% \textcolor{blue}{
Similarly, the connection and function of the internal node 15+$rc$ can be represented with [15+2$r$, 15, 2], because the addresses of its inputs are node 15+2$r$ and node 15, respectively, and AND is the 
% \textcolor{red}{\sout{second} 
logic type with index 2 in $\Gamma$.
The connection of each
% }
output node can be coded with 
% \textcolor{red}{\sout{one}
an
% }
integer, i.e., the address the output node connects to.
For example, the connection of the output node $O_{0}$ and $O_{m-1}$ can be represented with 16 and 15+2$r$, respectively.
In this way, any possible circuits represented by this 2D node array can be coded as a chain of integers, as shown at the bottom of \figname~\ref{fig_CGP}.
Any adjustment to this chain of integers is equivalent to the adjustment to an encoding.
% }

% \subsubsection{Determine the Hyperparameters}
% \textcolor{blue}{xxx}

\subsubsection{Search Algorithm for the Multiplier Design}
We adopt Cartesian Genetic Programming (CGP) \cite{miller1997designing,miller2015cartesian} to
% \textcolor{blue}{
adjust the chain of integers, a representation of the 2D node array, called a chromosome.
% }
CGP, a branch of genetic programming \cite{langdon2013foundations}, was initially developed for design and optimization of digital circuits.
With the CGP, chromosomes will be mutated and selected to minimize a cost function, so that a high quality chromosome can be obtained at the end of the algorithm.
% \textcolor{blue}{
We use two metrics to evaluate the quality of a chromosome: approximation error and area of 
% \textcolor{red}{\sout{an encoding-based multiplier}
the circuit candidate.
% }
To apply the CGP algorithm, the following items should be clearly defined:
a) cost function;
b) approximation error;
c) position weights.

\begin{itemize}[leftmargin=12pt]

\item[a)] \textit{Cost Function:}
To search a better chromosome, a cost function, denoted as $\mathcal{C}$, should be defined to evaluate the qualities of chromosomes.
% \textcolor{blue}{
The smaller value of cost means the higher quality of the chromosome.
% }
Since it is challenging to achieve minimum approximation error and minimum area overhead simultaneously,
% \textcolor{blue}{\sout{we partition the CGP search into two phases}
we define the cost function as a piecewise function, which consists of two sub-functions.
% }
The first
% \textcolor{blue}{\sout{phase targets}
sub-function guides the CGP algorithm
% }
to minimize the approximation error to meet a given threshold and the second
% \textcolor{blue}{\sout{phase is}
sub-function guides the CGP algorithm
% }
to minimize the area while maintaining the approximation error under the error threshold. 
An illustration of the effect of the two sub-functions during the CGP search process 
will be described in Section \ref{sec:Illustration of CGP Search Process} and shown in \figname~\ref{figCGP_search_process}. 
Accordingly, the cost function can be expressed as follows 
\begin{equation}\label{eq:cost_func}
\mathcal{C}=\begin{cases}
  \varepsilon  +  \mathcal{A}_{max},          &   \text{ if }~   \varepsilon   >    \varepsilon_{th} \\
  \varepsilon_{th}  +  \mathcal{A}_{\mathcal{M}^{\prime}},   &   \text{ if }~   \varepsilon   \le  \varepsilon_{th}
\end{cases},
\end{equation}
where $\varepsilon$ represents the approximation error, which will be defined
% \textcolor{blue}{\sout{later}
in Equation (\ref{eq:max_re}).
% }
$\varepsilon_{th}$ is a given threshold of approximation error. 
% , and we define $\varepsilon_{th}$ $\in$ $\{$$0.1\%$, $0.2\%$, $0.5\%$, $1\%$, $1.5\%$, $2\%$, $5\%$, $10\%$, $20\%$$\}$. 
% We did not employ $\varepsilon_{th}$ beyond 20\% since the accuracy of neural networks drops significantly. 
$\mathcal{A}_{\mathcal{M}^{\prime}}$ is the total area of the multiplier candidate evaluated with 15 nm NanGate cell library \cite{OpenCellLibrary15nm}. $\mathcal{A}_{max}$ is defined as the maximum area of all possible candidate multipliers, $\mathcal{A}_{max}=r \times c \times \text{area}_{max}$, where $\text{area}_{max}$ is the maximum area value of gates in $\Gamma$.
$\mathcal{A}_{max}$ is employed to balance cost values across two
% \textcolor{blue}{\sout{phases}
sub-functions,
% }
ensuring the second
% \textcolor{blue}{\sout{phase's}
sub-function's
% }
cost is definitely lower than the first.

\item[b)] \textit{Approximation Error:}
We define the approximation error of the multiplier as the largest normalized difference between the original multiplication result and the result represented with the new encoding, i.e., maximal relative error \cite{mrazek2016design}. 
% To calculate maximal relative error $\varepsilon$, we need to enumerate all possible inputs, compute the corresponding outputs, and compare them with those from an exact multiplier. 
% = \left \{ I_0, I_1, \cdots, I_i, \cdots,I_{2^n-1} \right \}
%Let $B^n=\left \{ 0,1 \right \}^n$ represent the set of $n$-bit Boolean vectors. 
Let $x$ and $y$ be $8$-bit Boolean vectors,  $z$ be a $16$-bit Boolean vector and $z^\prime$ be a $m$-bit Boolean vector. 
%$\left[b_0,b_1,\cdots,b_i,\cdots,b_{n-1}\right]$, where $b_i\in\left \{ 0,1 \right \}$.
Assume the original 8-bit accurate signed multiplier is implemented by a function $\mathcal{M}(x,y)\to z: B^8 \times B^8\to B^{16}$, 
where $B^8$ and $B^{16}$ are the 8- and 16-dimensional Boolean spaces, repectively.
The actual integer value represented by the 16-bit output $z$ of the multiplier in
two's complement can be expressed by a function $\delta(z): B^{16}\to \mathbb{N}$, as
$\delta(z)=-2^{15}b_{15} + \sum_{i=0}^{14} 2^{i}b_{i}$, where $b_0\dots b_{15}$ are the individual bits in $z$.
Similarly, let a candidate of the proposed 8-bit encoding-based signed multiplier with $m$ output nodes be represented by a function $\mathcal{M}^{\prime}(x,y)\to z^\prime: B^8 \times B^8\to B^{m}$, 
and the value of the $m$-bit output $z^\prime$
be expressed as
 $\delta^{\prime}(z^\prime): B^{m}\to \mathbb{N}$ in the proposed new encoding, i.e., 
$\delta^{\prime}(z^\prime)=\sum_{i=0}^{m-1} s_ib_i$, where $\textbf{\textit{s}}=\left [ s_0, s_1, \dots, s_{m-1}\right ] \in \mathbb{N}^m$ is the set of the position weights. 
Accordingly, the maximal relative error $\varepsilon$ can be defined as follows
\begin{equation}\label{eq:max_re}
%\forall_{(a,b) \in B^N \times B^N} : 
\varepsilon = \max_{\forall \left(x, y\right)\in B^8\times B^8}\frac{\left | \delta \left ( \mathcal{M} \left(x, y\right) \right )   - \delta^{\prime} \left ( \mathcal{M}^{\prime} \left(x, y\right)  \right ) \right | }{  \max \left | \delta \left (  \mathcal{M}(x,y)    \right ) \right |
},
\end{equation}
where $\max \left | \delta \left (  \mathcal{M}(x,y)  \right ) \right | $ is the maximum absolute value of the output of the original multiplier.

\item[c)] \textit{Position Weights:}
The position weights $\textbf{\textit{s}}$ in $\delta^{\prime}$
% \textcolor{blue}{\sout{is}
are
% }
determined as follows
\begin{equation}\label{eq:lst_PositionWeights}
 \textbf{\textit{s}}^{*} = 
  \mathop{\arg\min}\limits_{\textbf{\textit{s}}\in \mathbb{N}^m} \sum_{\forall \left(x, y\right)\in B^8\times B^8}
  {\left | \delta \left ( \mathcal{M}  \left(x, y\right)  \right )  - \delta^{\prime} \left ( \mathcal{M}^{\prime}  \left(x, y\right) \right ) \right |^2}.
\end{equation}

\ \ \ To solve this optimization problem, we enumerate all the 65,536 possible input combinations of $x$ and $y$. 
Assume $x_0\dots x_{255}$ are the 256 combinations of $x$ and $y_0\dots y_{255}$ the 256 combinations of $y$.    
Let $\textbf{\textit{v}}$ represent a column vector of the integer values of the
original multiplier output, and
 $\textbf{\textit{B}}$ represent a matrix of candidate multiplier output bits
 as follows
\begin{equation}\label{eq:value_B}
\footnotesize
 \textbf{\textit{v}} = 
\begin{bmatrix}
 \delta \left ( \mathcal{M}  \left(x_0, y_0\right)  \right )\\
 \delta \left ( \mathcal{M}  \left(x_0, y_1\right)  \right )\\
\vdots \\
 \delta \left ( \mathcal{M}  \left(x_0, y_{255}\right)  \right )\\
 \delta \left ( \mathcal{M}  \left(x_1, y_0\right)  \right )\\
\vdots \\
 \delta \left ( \mathcal{M}  \left(x_1, y_{255}\right)  \right )\\
\vdots \\
 \delta \left ( \mathcal{M}  \left(x_{255}, y_{255}\right)  \right )\\
\end{bmatrix}\in \mathbb{N}^{2^{16}},
\textbf{\textit{B}} = 
\begin{bmatrix}
 \mathcal{M}^{\prime}  \left(x_0, y_0\right)\\
 \mathcal{M}^{\prime}  \left(x_0, y_1\right)\\
\vdots \\
 \mathcal{M}^{\prime}  \left(x_0, y_{255}\right)\\
 \mathcal{M}^{\prime}  \left(x_1, y_0\right)\\
\vdots \\
 \mathcal{M}^{\prime}  \left(x_1, y_{255}\right)\\
\vdots \\
 \mathcal{M}^{\prime}  \left(x_{255}, y_{255}\right)\\
\end{bmatrix}\in B^{2^{16}\times m}.
\end{equation}
where $ \mathcal{M}^{\prime}$ outputs
% \textcolor{blue}{\sout{a}
an
% }
$m$-bit Boolean vector at the output of the proposed multiplier.

\ \ \ \ Assume the position weights to be determined are written as $\textbf{\textit{s}}=\left [ s_0, s_1, \dots, s_{m-1}\right ]^T \in \mathbb{N}^{m}$. 
Equation (\ref{eq:lst_PositionWeights}) can then be rewritten as follows
\begin{equation}\label{eq:lst_PositionWeights_v2}
 \textbf{\textit{s}}^{*} = 
\mathop{\arg\min}\limits_{\textbf{\textit{s}}\in \mathbb{N}^{m}} 
  \left \| \textbf{\textit{v}} - \textbf{\textit{Bs}} \right \|_2.
\end{equation}

\ \ The column vectors of matrix \textbf{\textit{B}} correspond one-to-one with the output nodes. 
% \textcolor{blue}{\sout{From a statistical perspective, t}T}
The column vectors of \textbf{\textit{B}} serve as the independent variables, and the vector \textbf{\textit{v}} as the dependent variable, making the process of determining position weights akin to solving for coefficients in a linear regression model. 
However, since the candidate multipliers are generated by the CGP algorithm, it cannot be guaranteed that the column vectors of \textbf{\textit{B}} are linearly independent, potentially leading to multicollinearity issues \cite{dormann2013collinearity,daoud2017multicollinearity}. 
So, multicollinearity in our case refers to a situation in which two or more column vectors of \textbf{\textit{B}} are highly linearly correlated, implying that one column can be accurately predicted from a combination of others. 
To address this issue, we use ridge regression \cite{hoerl1970ridge} to determine the position weights as follows
\begin{equation}\label{eq:lst_PositionWeights_v3}
 \textbf{\textit{s}}^{*} = \text{round}\left ( \left ( \textbf{\textit{B}}^T\textbf{\textit{B}} + \lambda \textbf{\textit{I}} \right )^{-1}\textbf{\textit{B}}^T \textbf{\textit{v}} \right ),
\end{equation}
where \textbf{\textit{I}} is an identity matrix, and $\lambda$ is a regularization parameter. 
We set $\lambda=0.1$. Rounding is applied as the position weights will be represented with signed integers. 

\end{itemize}

% \begin{equation}\label{eq:cost_func}
% \mathcal{C} =\left\{\begin{matrix} 
%   \text{\textcolor{blue}{\sout{-(}}}~~\varepsilon ~+~  \mathcal{A}_{max}\text{\textcolor{blue}{\sout{)}}},  \quad \quad~~~\text{if} \quad \varepsilon > \varepsilon_{th} 
% \\  
%   \text{\textcolor{blue}{\sout{-(}}}\varepsilon_{th} +~ \mathcal{A}_{\mathcal{M}^{\prime}}\text{\textcolor{blue}{\sout{)}}},  \quad \quad~~~~\text{if} \quad \varepsilon \le  \varepsilon_{th} 
% \end{matrix}\right. ,
% \end{equation}

\smallskip

The complete search process includes two parts: a) simple grid search to determine the hyperparameters of CGP; b) CGP search sweep across different output bit-width $M$ and logic level $c$ for each maximal relative error threshold $\varepsilon_{th}$.

\begin{itemize}[leftmargin=12pt]

\item[a)] \textit{Simple Grid Search:}\label{sec:grid_search}
The simple grid search is shown in Algorithm (\ref{Alg:grid_search}). 
The hyperparameters of CGP include the number of logic levels $c$, logic depth $r$, and the number of output nodes $m$.
To reduce the design space complexity, we limit the number of logic levels to a maximum of three (line 2). 
The objective of the grid search is to identify the optimal logic depth and number of output nodes for various logic levels. 
Thus, the grid search is across different values of two hyperparameters: $r$ and $m$. For each hyperparameter, three different values are considered, namely 64, 128 and 256 (line 5 to line 6). 
% We did not employ the value beyond 256 as it would make the search process highly time-consuming. 
So a total of nine different combinations are evaluated and compared for each logic level $c$.
The output bit-width $M$ is fixed at 64 bits,
% \textcolor{blue}{
because 64-bit is enough to reach a very low maximal relative error.
% }
To guide the CGP algorithm to minimize the maximal relative error as much as possible, the maximal relative error threshold $\varepsilon_{th}$ is set to $0.0\%$. For each logic level $c$, the combination that achieves the smallest maximal relative error will be selected as the optimal hyperparameter setting (line 9 to line 12).
The optimal hyperparameter settings are then used for the subsequent CGP search sweep in Algorithm (\ref{Alg:CGP_search}).

\begin{algorithm}[t]
\caption{Grid Search for Hyperparameters of CGP} % to determine the minimal bit-width and corresponding encoding system for multiplier output}
\footnotesize
\begin{algorithmic}[1]
\Statex \textbf{Input:} maximal relative error threshold $\varepsilon_{th} = 0.0\%$, output bit-width 
\Statex \ \ \ \ $M=64$

\Statex \textbf{Output:} the best hyperparameters configuration for each logic level
% logic depth $r$ and output nodes $m$ for each logic level $c$

\Statex 
\State $configs \gets [~]$
\For{logic level $c \gets \{1, 2, 3\}$}
\State $cost_{\text{best}} \gets \infty$
\State $config_{\text{best}} \gets \text{None}$
\For{logic depth $r \gets \{64, 128, 256\}$}

\For{output nodes $m \gets \{64, 128, 256\}$}
\State Based on given $r$, $c$, $m$, $M$, $\varepsilon_{th}$, run Algorithm (\ref{Alg:CGP_search})
\State Save $cost$ returned from Algorithm (\ref{Alg:CGP_search})
\If{$cost < cost_{\text{best}}$}
\State $cost_{\text{best}} \gets cost$
\State $config_{\text{best}} \gets (r, m)$
\EndIf
\EndFor

\EndFor
\State Append $config_{\text{best}}$ to $configs$ for logic level $c$
\EndFor
\State \Return $configs$
\end{algorithmic}
\label{Alg:grid_search}
\end{algorithm}

\begin{algorithm}[t]
\caption{CGP Search for
% \textcolor{red}{\sout{Encoding-based}
Multiplier Design
% }
} % to determine the minimal bit-width and corresponding encoding system for multiplier output}
\footnotesize
\begin{algorithmic}[1]
\Statex \textbf{Input:} logic depth $r$, logic level $c$, output nodes $m$, output bit-width $M$,
\Statex \ \ \ \ maximal relative error threshold $\varepsilon_{th}$
%  \in \{0.1\%, 0.2\%, 0.5\%, 1\%, 1.5\%, 2\%, 5\%, 10\%, 20\%\}

\Statex \textbf{Output:} the best individual

\Statex 
\State input nodes $n=$16
\State $n_{\text{parents}}=10$, $n_{\text{offsprings}}=50$, $n_{\text{champions}}=2$, $n_{\text{generations}}=2500$
\State gate library $\Gamma=\{$$0^{identity}$, $1^{not}$, $2^{and}$, $3^{or}$, $4^{xor}$, $5^{nand}$, $6^{nor}$, $7^{xnor}$, $8^{const0}$, $9^{const1}$$\}$
% \State // generate initial population
\For{$i \gets 1 : (n_{\text{parents}} + n_{\text{offsprings}})$}
\State Randomly generate individual $i$
\State Evaluate individual $i$ with OBJECTIVE function
\EndFor
% \State // determine the parents
\State Select the top $n_{\text{parents}}$ individuals with the lowest cost as the parents
% \State // evolution starts
\For{$g \gets 1 : n_{\text{generations}}$}
% \State // mutation
\For{$i \gets 1 : n_{\text{offsprings}}$}
\State Mutate the parents to generate offspring $i$
\State Evaluate offspring $i$ with OBJECTIVE function 
\EndFor
\State Select the top $n_{\text{champions}}$ offspring with the lowest cost as the champions
\State Update the parents with the champions according to the rules in \cite{miller2020cartesian}
% \For{$i$ in range($n_{\text{champions}}$)}
% \If{Champion $i$ has \textit{better} fitness than any parents}
% \State Replace the worst parent with the champion
% \ElsIf{Champion $i$ has \textit{equal} fitness to the parent}
% \State Randomly select one of these as parent
% \Else
% \State Keep the parents unchanged
% \EndIf
% \EndFor

\EndFor
% \State Identify the best individual in the parents
\State \Return the best individual in the parents

\Statex 
% \Statex // define fitness function
\Function{\normalsize objective}{individual, $n$, $m$, $M$, $\varepsilon_{th}$}
% \State For all possible inputs, calculate corresponding outputs to obtain a truth table
% \State For $\forall (I_1,I_2)\in \{0,1\}^8\times\{0,1\}^8$, calculate $\mathcal{M}(I_1,I_2)$ and $\mathcal{M}^{\prime}(I_1,I_2)$
% \State Convert the outputs to a 2D matrix \textbf{\textit{B}} with $2^n$ rows and $m$ columns
% \State \textbf{\textit{B}} = maximal set of linearly independent columns from \textbf{\textit{B}}
\State Enumerate all possible inputs, construct \textbf{\textit{v}} and \textbf{\textit{B}} with Equation (\ref{eq:value_B})
\State Determine position weights with Equation (\ref{eq:lst_PositionWeights_v3})
\State Identify the \textit{indexes} of the $M$ largest absolute values in position weights
\State Select the output nodes with \textit{indexes} as the $M$-bit output of candidate multiplier
% \State Construct candidate multiplier $\mathcal{M}^{\prime}$
% \State Create a $M$-bit output candidate multiplier with the \textit{index}
% \State Calculate maximal relative error $\varepsilon$ with Equation ()
% \State Calculate area $\mathcal{A}_{\mathcal{M}^{\prime}}$
\State Calculate maximal relative error $\varepsilon$ with Equation (\ref{eq:max_re}) and area $\mathcal{A}_{\mathcal{M}^{\prime}}$
% \State Calculate fitness with Equation ()
\State \Return the cost determined by Equation (\ref{eq:cost_func})
\EndFunction

\end{algorithmic}
\label{Alg:CGP_search}
\end{algorithm}

\item[b)] \textit{CGP Search Sweep:}\label{sec:cgp}
The CGP search sweep is across all the possible combinations of output bit-width $M \in \{ 36$, $40$, $42$, $44$, $46$, $48$, $52$, $56$, $60$, $64\}$ and logic level $c\in\{1,2,3\}$ for each maximal relative error threshold $\varepsilon_{th}$ $\in$ $\{$$0.1\%$, $0.2\%$, $0.5\%$, $1\%$, $1.5\%$, $2\%$, $5\%$, $10\%$, $20\%$$\}$, respectively.
For each combination, the CGP search process is presented in Algorithm (\ref{Alg:CGP_search}) and includes the following steps: 
First, some local parameters are defined (line 1 to line 3). 
Second, 60 individuals are randomly generated and each individual is evaluated using the OBJECTIVE function (line 4 to line 7). 
Third, the 10 individuals with the lowest cost are selected as the initial parents (line 8). 
In each generation, the parents are mutated to generate 50 offsprings and each offspring is assessed using the OBJECTIVE function (line 10 to line 13). 
Then, the two offsprings with the optimal cost are selected as champions to update the parents (line 14 to line 15). 
The updated parents will be used in the next generation. 
The mutation rate for each generation is dynamically adjusted based on the maximal relative error of the previous generation's champions. 
The algorithm terminates
% \textcolor{blue}{\sout{until}
when
% }
it reaches the maximum number of generations (line 17).

\ \ \ Due to the huge search space, we deliberately set the number of output nodes $m$ to exceed the output bit-width $M$ to expand the exploration space. 
The increased number of output nodes provides more possibilities for circuit design. 
The algorithm evaluates the importance of each output node and consistently selects only the most significant nodes as the final output bits to construct a candidate multiplier.
Therefore, the OBJECTIVE function in Algorithm (\ref{Alg:CGP_search}) includes three parts: 
evaluating the importance of each output node (line 19 to line 21), 
selecting important output nodes to construct a candidate multiplier (line 22), 
and calculating the cost (line 23 to line 24). 
From Equation (\ref{eq:lst_PositionWeights_v2}) and (\ref{eq:lst_PositionWeights_v3}), each output node is associated with a position weight. 
Therefore, the importance of each output node is identified by the absolute value of its position weight (line 21).

\ \ \ 
% \textcolor{blue}{
Two examples of candidate multipliers in CGP search sweep are shown in \figname~\ref{fig:logicLibSample}, where $r=256$, $c=1$, $m=64$, $M=48$, and $\varepsilon_{th}$ for (a) and (b) is 5\% and 1\%, respectively. The position weights of output bits are shown at the right side of the outputs.
At the end of each CGP search, the final searched circuit will be used to constructed an encoding-based multiplier.
% }

\end{itemize}

% \subsubsection{Simple Grid Search}\label{sec:grid_search}

% \subsubsection{CGP Search Sweep}\label{sec:cgp}

\begin{figure}[t]
    % \centering
    % \centerline{\includegraphics[width=0.26\linewidth]{./Figs/logic_sample.pdf}}
    % \caption{One sample of logic mapping from input bits to output bits of a multiplier. }
    % \label{fig:logicLibSample}
% \subfigure[]{
% \begin{minipage}[c]{0.31\linewidth}
% \centering
% % \label{fig:mul-trad}
% \includegraphics[width=1\textwidth]{Figs/logic_samples_1.pdf}
% \end{minipage}%
% }%
% \hspace{0.01\linewidth}
\subfigure[]{
% \subfloat[]{
\begin{minipage}[c]{0.45\linewidth}
\centering
% \label{fig:mul-trad}
\includegraphics[width=1\linewidth]{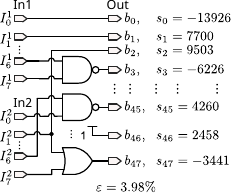}
\end{minipage}%
}%
\hspace{0.04\linewidth}
\subfigure[]{
% \subfloat[]{
\begin{minipage}[c]{0.45\linewidth}
\centering
% \label{fig:mul-trad}
\includegraphics[width=1\linewidth]{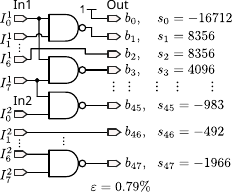}
\end{minipage}%
}%
\caption{
Two examples of logic mapping from input bits to output bits of a candidate multiplier. The position weights evaluated in each example are shown at the outputs. The resulting maximal relative error of each candidate is illustrated at the bottom. (a) A circuit candidate with a large maximal relative error. (b) A circuit candidate with a small maximal relative error.
}
\label{fig:logicLibSample}
\end{figure}

\subsection{Adder Design with Encoding}\label{sec:Methodology-ADD}
\begin{figure}[t]
\centerline{\includegraphics[width=1\linewidth]{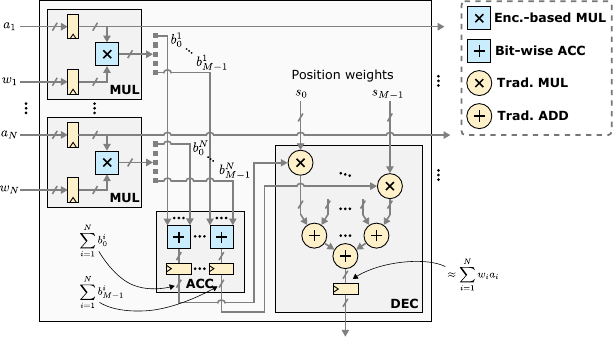}}
\caption{ A column in a MAC array consists of encoding-based multipliers and the circuit of the addition function, which consists of bit-wise accumulator (ACC) and a decoder (DEC).
%The circuit implementing the addition function and the MAC array.  %acomparison between traditional design in binary system (two's complement) and an example of proposed design in our encoding. (a) MAC in binary. (b) MAC in our encoding.
%Illustration of proposed binary search algorithm for the encoding of the product of 8-bit uniform quantization levels. Left: RMSE vs. iterations of search algorithm. Right: RMSE vs. the number of random samples.
}
\label{fig:ourMAC}
\end{figure}

In the section above, we have determined the structure of a multiplier using a new encoding. For a column of $N$ multipliers in a MAC array, e.g, \figname~\ref{fig:ourMAC},
%Since the bit sequences at the output of the multiplier do not follow the two's complement number system, we also need to define a new structure to implement the addition function.  
%For the general case of accumulating the $M$-bit outputs of $N$ multipliers, 
the sum can be expressed as $\sum_{i=1}^{N}\sum_{j=0}^{M-1}s_j\times b_j^i=\sum_{j=0}^{M-1}s_j\times \sum_{i=1}^{N}b_j^i$, where $b_j^i$ is the $j$-th bit of the output of the $i$-th multiplier. Accordingly, the circuit to implement this sum can be designed as illustrated in \figname~\ref{fig:ourMAC}. In this design, the same bits  at the outputs of the encoding-based multipliers in the same column of a MAC array are added together with the ACC units. For the multipliers with $M$ output bits, $M$ ACC units are needed.
Since these bits are not weighted, the implementation of such ACC units is very simple.
In the DEC unit, the outputs of ACC units are multiplied with the position weights, and the results are then accumulated by an adder tree 
to calculate the result of a whole column in the
% in the MAC array
two's complement format. These position weigths are 
% determined to approximate the results of multipliers, 
% they are 
identical and unchanged for all the neural networks, and thus fixed
% in the circuit
to simplify the logic of the mutipliers and the adder tree
in the 
% used to multiply the position weights with the outputs of
DEC unit.

\subsection{Design and Application of MAC Array}\label{sec:Methodology-MAC}

We deploy the encoding-based multipliers and adders to construct a MAC array with a size $N \times N$ to  execute MAC operations in DNNs efficiently, one column of which is illustrated in \figname~\ref{fig:ourMAC}. 
At the inputs of each encoding-based multiplier, flip-flops are inserted to allow the reuse of inputs, similar to that in the traditional systolic array. 
% \textcolor{blue}{\sout{While compared}
Compared
% }
to the traditional systolic array, our MAC array design differs in two aspects: 
First, whereas each MAC unit in the traditional systolic array includes an individual adder, in our design, adders are only placed at the bottom of each column, namely ACC units, which performs the bit-wise accumulation in a column. 
Second, traditional systolic arrays require flip-flops in each MAC unit to store partial sums. 
Due to the short critical paths inside the encoding-based multipliers and highly parallelized bit-wise accumulators, flip-flops
% \textcolor{blue}{\sout{are}}
only appear at the outputs of the ACC units in our design. 
Furthermore, we can utilize the \texttt{optimize\_registers} command in Design Compiler to optimize the number and placement of these registers during the logic synthesis process to enhance the area and power efficiency.

To execute MAC operations with the encoding-based MAC array, weights in a neural network are first loaded into the flip-flops in each multiplier.
Activations are streamed as inputs. 
Unlike traditional MAC units in systolic array that require flip-flops for storing intermediate partial sums, 
activations in our design are not required to be shifted as in the traditional systolic array. 
Activations belonging to the inputs of a neuron can enter each column simultaneously and the results of multiplication and bit-wise accumulation are obtained after two clock cycles.

The encoding-based MAC array
% \textcolor{blue}{\sout{have}
has
% }
slightly better throughput and latency
% \textcolor{blue}{\sout{than those of}
compared to
% }
the traditional systolic array under the same size while achieving a much lower area cost and power consumption. Assume that a weight matrix
% $W$
with a size of $N \times N$ has been loaded into the encoding-based array and the traditional MAC array
% \textcolor{blue}{
with a same size of $N \times N$.
% }
The clock period is denoted as $T$. To finish  a computation between an input matrix
% $I_0$
with a size of $N \times N$, the latency of the encoding-based array and the traditional MAC array are $2NT$ and $(3N-2)T$, respectively. %When $N$ is large enough, their latencies are nearly the same \textcolor{red}{(need discussion)}. 
To evaluate the throughput, we assume that
% \textcolor{blue}{\sout{$m$}
$l$
% }
input matrices with sizes of $N \times N$ need to be processed by the MAC arrays. The throughputs of the encoding-based array and the traditional MAC array are
$l / ((2N + N(l-1))T)$
% $\frac{m}{[(2N-1) + N(m-1)] \times T}$
and
$l / ((3N-2 + N(l-1))T)$
% $\frac{m}{[(3N-2) + N(m-1)] \times T}$
, respectively.
The proposed design exhibits a
% \textcolor{blue}{\sout{higher performance}
slightly smaller latency and higher throughput,
% }
and the throughputs of these designs become nearly the same when $l$ goes larger
% \textcolor{blue}{
and larger.

\subsection{Fine-tune Neural Networks}\label{sec:finetune} 
% \textcolor{blue}{\sout{To verify the performance of proposed encoding-based MAC array, we integrate the proposed encoding-based multiplier into neural networks to test for accuracy.}}

To test the accuracy of neural networks, 
we modify the \textit{Linear} and \textit{Conv2d} classes in PyTorch by replacing the function of the original 8-bit multiplier with
% \textcolor{blue}{\sout{a Look-Up Table (LUT)}
a searched encoding-based multiplier.
% }
% \textcolor{blue}{\sout{The LUT maps the inputs to their corresponding outputs of the proposed encoding-based multiplier.}} 
The position weights
% \textcolor{blue}{
in all modified \textit{Linear} and \textit{Conv2d} layers
% }
are identical.
% \textcolor{blue}{\sout{in all LUTs of all \textit{Linear} and \textit{Conv2d} layers}}. 
Due to the error incurred by multiplier approximation, the accuracy of neural networks degrades. 
To address this issue, the trainable parameters of neural networks will be fine-tuned to compensate the accuracy degradation. 
The position weights are unchanged during fine-tuning.
% , and thus can be used to simplify the multipliers and adder tree after ACC units. 
During forward propagation, exact multiplications are replaced with approximate operations as follows
\begin{equation}\label{eq:finetune_forward}
\text{Forward: }   p = f(a, b)=g(q), q=ab,
\end{equation}
% where $a$, $w$, and $p$ stands for \textit{activation}, \textit{weight}, and \textit{product}, respectively. 
where function $f$ denotes the encoding-based multiplier that maps two numbers, $a$ and $b$, to one number, $p$. 
Function $g$ maps the exact multiplier output $q$ to the approximate counterpart $p$. 
Replacing exact multiplication with approximate operations during forward propagation renders backpropagation non-differentiable.
However, function $g$ works like a quantization operation to the exact multiplication results of $a$ and $b$. 
So, we can use the straight-through estimator (STE) \cite{bengio2013estimating} to estimate the derivative of function $g$, namely $\frac{\mathrm{d} p}{\mathrm{d} q}\approx1$. Thus, the partial derivatives of $p$ with respect to $a$ and $b$ are demonstrated as follows
\begin{align}\label{eq:finetune_backward}
\text{Backward: }
\left\{
	\begin{aligned}
         \frac{\partial p}{\partial a} = \frac{\mathrm{d} p}{\mathrm{d} q} \frac{\partial q}{\partial a} \approx \frac{\partial q}{\partial a}=b \\
         \frac{\partial p}{\partial b} = \frac{\mathrm{d} p}{\mathrm{d} q} \frac{\partial q}{\partial b} \approx \frac{\partial q}{\partial b}=a
	\end{aligned}
\right. .
\end{align}

\begin{figure}[t]
% \vspace{8pt}
\centerline{\includegraphics[width=0.781\linewidth]{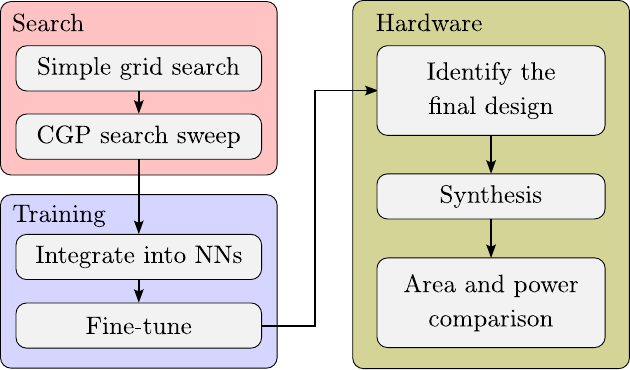}}
\caption{Experimental flowchart outlining the main steps of the experiments conducted in this study.}
\label{fig_experimental_flowchart}
\end{figure}

\section{Experimental Setup and Results}\label{sec:results} 
%\textcolor{red}{Throughput and latency are not mentioned in the exprimental results.}
\subsection{Experimental Setup}\label{sec:ex_setup}

\subsubsection{Setup for Hardware Synthesis}\label{Setup for Hardware Synthesis}
To verify the proposed encoding-based MAC array, we performed circuit synthesis with the NanGate 15 nm cell library \cite{OpenCellLibrary15nm}. 
An encoding-based multiplier was searched by the CGP algorithm and then used to construct 
% These MAC circuits approximate the results of the uniformly quantized 8-bit MAC units.  
% Such MAC circuits were then used to construct 
a MAC array 
%similar to the systolic array in Google's TPU \cite{TPU}. 
% \textcolor{blue}{\sout{similar}
according
% }
to \figname~\ref{fig:ourMAC}. 
It was sufficient to represent the position weights with 16-bit signed integers
% \textcolor{blue}{
in our searched encoding-based multipliers.
% }
% The position weights were sufficiently represented with 15-bit signed integers.
The position weights are
% \textcolor{blue}{
determined by the CGP algorithm and
% }
unchanged
% \textcolor{blue}{
during fine-tuning,
% }
and thus can be fixed to simplify the multipliers and adder tree after ACC units in \figname~\ref{fig:ourMAC}. 
In synthesizing the circuits, the clock frequency was set to 
% \textcolor{blue}{\sout{1 GHz}}
5 GHz, as this is the maximum clock frequency that traditional MAC units (\figname~\ref{fig:SystolicArray}(b)) can achieve with the NanGate 15 nm cell library \cite{OpenCellLibrary15nm}.
Power and area analysis
% \textcolor{blue}{\sout{of this hardware}}
were conducted with Design Compiler from Synopsys \cite{bhatnagar2007advanced}. 
The \texttt{optimize\_registers} command in Design Compiler was used to optimize the number and placement of the registers
after ACC units
% at the outputs of ACC units
to enhance the area and power efficiency.

% \textcolor{blue}{\sout{Such}
\subsubsection{Setup for Neural Network Inference and Fine-tuning}
To verify that such
% }
a MAC array can efficiently execute neural networks while maintaining a high inference accuracy,
% \textcolor{blue}{\sout{. To verify this}}
we tested the accuracy of six neural networks together with the corresponding datasets, ResNet18 (Cifar-10) \cite{He_2016_CVPR}, ResNet20 (Cifar-100), ResNet50 (ImageNet), 
% \textcolor{blue}{
MobileNetV2 (Cifar-10) \cite{Sandler_2018_CVPR}, MobileNetV2 (Cifar-100), and EfficientNet-B0 (ImageNet) \cite{pmlr-v97-tan19a}
% }
, 
%on the constructed hardware. 
using the new MAC design. 
8-bit uniform quantization was applied. 
The neural networks above were initially trained using PyTorch, and pretrained weights were loaded from Torchvision \cite{ImageNetpytorch} 
and public repositories on GitHub \cite{Cifar10pytorch, pytorchCifar100}. 
% For ResNet18 and ResNet20, Nvidia Quadro RTX 6000 GPU 24 GB was used for training, and for ResNet50 Nvidia A100 80GB GPU was used.
% The learning rate of the position weights in the novel encoding for fine-tuning ResNet18, ResNet20, and ResNet50 was set to 1e-3, 1e-3, and 1e-8, respectively.
% To apply our encoding-based multiplier, we override \textit{Linear} and \textit{Conv2d} class in PyTorch, and embed the encoding-based multiplier as a Look-Up Table (LUT) to replace the original exact 8-bit multiplier. 
% Due to the approximation in the proposed encoding technique, the accuracy of neural networks cannot be maintained initially. 
% To address this, we fine-tune the neural networks for 25 epochs to compensate the accuracy degradation. 
% The learning rate used to fine-tune ResNet18, ResNet20, and ResNet50 was set to 1e-4, 1e-4, and 1e-5, respectively.
% After overriding the \textit{Linear} and \textit{Conv2d} layer in PyTorch, the fine-tuning of ResNet50 on ImageNet becomes very time-comsuming. 
% To solve this problem, in every epoch, we only randomly sample 1\% images of each class in original training dataset for fine-tuning, while execute the inference on all testing dataset.
% To integrate the proposed encoding-based multiplier into the new MAC array for fine-tuning and inference, 
% we modified the \textit{Linear} and \textit{Conv2d} classes in PyTorch by 
% replacing the function of the original 8-bit multiplier with a Look-Up Table (LUT). Due to the inaccuracy incurred by multiplier approximation, the accuracy of neural networks degrades. 
% To address this issue, 
Then, the neural networks were fine-tuned 
% \textcolor{blue}{
on an NVIDIA A100 80 GB
% } 
for 25 epochs, in which  % to integrate our encoding-based multiplier, implementing it as a Look-Up Table (LUT).  to substitute the original precise 8-bit multiplier. 
%Due to the inherent approximation of our proposed encoding technique, the initial accuracy of the neural networks was impacted. 
%To mitigate this accuracy loss, we fine-tuned the neural networks for 25 epochs. 
%Specifically, 
the learning rates %for fine-tuning 
% \textcolor{blue}{
for MobileNetV2 (Cifar-100) were set at 1e-3, 
for ResNet18 (Cifar-10), ResNet20 (Cifar-100) and MobileNetV2 (Cifar-10) at 1e-4, and 
for ResNet50 (ImageNet) and EfficientNet-B0 (ImageNet) at 1e-5. 
% }
After rewriting the \textit{Linear} and \textit{Conv2d} layers, finetuning ResNet-50 and EfficientNet-B0 on ImageNet becames extremely time-consuming. 
To address this issue, 
% To speed up the fine-tuning ResNet-50 on ImageNet, 
%during each finetuning epoch, 
we only randomly sampled 1\% of the images from each category in the original training dataset during each epoch. %, while the inference was still carried out on the entire test dataset.
% \textcolor{blue}{
For comparison, neural networks with 8-bit uniform quantization and 8-bit exact multipliers were also fine-tuned with 25 epoches.
% }
% \textcolor{red}{
% In fine-tuning, straight-through estimator (STE) \cite{bengio2013estimating} was used for propagating gradients of encoding-based multipliers. 
% }

\begin{table}[t]
\setlength{\tabcolsep}{2.4pt}%2.5pt
\renewcommand{\arraystretch}{1.0}
\centering
\footnotesize
\caption{Power and area of proposed vs. traditional MAC arrays. 
% \textcolor{blue}{5 GHz}
}
\begin{tabular}{ccccccccccc}

\toprule
\multirow{3}{*}{\begin{tabular}[c]{@{}c@{}}Size of\\MAC \\Array \end{tabular}} &
\multicolumn{2}{c}{\begin{tabular}[c]{@{}c@{}}Bit-Wid. of \\ Product\end{tabular}} & &
  \multicolumn{3}{c}{\begin{tabular}[c]{@{}c@{}}Area (mm$^2$)\end{tabular}} & &
  \multicolumn{3}{c}{\begin{tabular}[c]{@{}c@{}}Power (W)\end{tabular}} \\ \cmidrule{2-3} \cmidrule{5-7} \cmidrule{9-11}
 &
 \multicolumn{1}{c}{Trad.} &
  \multicolumn{1}{c}{Prop.} &
   &
  \multicolumn{1}{c}{Trad.} &
  \multicolumn{1}{c}{Prop.} &
  Red. &
   &
  \multicolumn{1}{c}{Trad.} &
  \multicolumn{1}{c}{Prop.} &
  Red. \\  
  \midrule
64$\times$64   &16&64&& \multicolumn{1}{c}{ 0.756} & \multicolumn{1}{c}{0.577} & 23.69\% & & \multicolumn{1}{c}{ 3.816} & \multicolumn{1}{c}{ 2.171} &  43.10\% \\ 
128$\times$128 &16&64&& \multicolumn{1}{c}{ 3.092} & \multicolumn{1}{c}{2.148} & 30.53\% & & \multicolumn{1}{c}{15.462} & \multicolumn{1}{c}{ 7.993} &  48.31\% \\ 
256$\times$256 &16&64&& \multicolumn{1}{c}{12.837} & \multicolumn{1}{c}{6.574} & 48.79\% & & \multicolumn{1}{c}{65.647} & \multicolumn{1}{c}{23.362} &  64.41\% \\ 
\bottomrule
\end{tabular}
\label{tab:generalHardwareResults}
\end{table}

\subsubsection{Setup for CGP search}
As mentioned in Section \ref{sec:grid_search} and Algorithm (\ref{Alg:grid_search}), a simple grid search was applied to determine the hyperparameters of CGP search algorithm, i.e., the configuration of logic depths $r$ and output nodes $m$ for each logic level $c$. 
The optimal configurations were then used for the subsequent CGP search sweep. 
% \textcolor{blue}{
In the experiments,
we ran Algorighm (\ref{Alg:grid_search}) on 15 servers
with Xeon E-2124G 3.40 GHz 32 GB RAM and Quadro RTX 6000 24 GB GPU cards. The total runtime was around 20.5 hours. Since this search process needs to be
executed only once and offline, the runtime is still reasonable in the proposed context.
% }
The CGP search algorithm swept across all the possible combinations of output bit-width $M \in \{ 36$, $40$, $42$, $44$, $46$, $48$, $52$, $56$, $60$, $64\}$ and logic level $c\in\{1,2,3\}$ for each maximal relative error threshold $\varepsilon_{th}$ $\in$ $\{$$0.1\%$, $0.2\%$, $0.5\%$, $1\%$, $1.5\%$, $2\%$, $5\%$, $10\%$, $20\%$$\}$, respectively. 
We did not apply $M$ beyond 64 as the accuracy of neural networks can be well maintained. 
We did not employ $\varepsilon_{th}$ beyond 20\% since the accuracy drops significantly. 
For each threshold $\varepsilon_{th}$, we selected the multiplier with the smallest area from the search results of all combinations of different output bit-width and logic levels as the best multiplier for that threshold. 
We integrated the best multiplier for each threshold into the neural networks, and tested its inference accuracy after fine-tuning. 
Among the multipliers that reach a minimal accuracy loss, we selected the one with the highest area and power efficiency as our final design to build the MAC array for area and power comparison.
% \textcolor{blue}{\sout{t}d the}

% \textcolor{blue}{
An experimental flowchart outlining the main experiments conducted in this paper is shown in \figname~\ref{fig_experimental_flowchart}. 
% \textcolor{blue}{
The open source code of this work can be found on GitHub with link
``\url{https://github.com/Bo-Liu-TUM/EncodingNet/}''.

\subsection{Experimental Results}\label{sec:ex_results}
%To verify the proposed method, we first conducted a generalized weighted encoding search 
%for the product of 8-bit uniform quantization levels. 
%Based on the searched encoding scheme, 
%the corresponding netlist description of general-purpose systolic array was generated and 
%synthesized with the NanGate 15 nm cell libraries \cite{OpenCellLibrary15nm} and the clock frequency around 1 GHz. 
%Power and area analysis were conducted in Design Compiler from Synopsys.

% \subsubsection{Sy}
\subsubsection{Hardware Synthesis of Encoding-based MAC Array}
Table \ref{tab:generalHardwareResults} shows the comparison between the proposed MAC array and the traditional systolic array in power consumption and area cost. Different sizes were used to verify the advantages of the hardware platform, as shown in the first column. The second and the third columns are the bit width of product, i.e., multiplication result, in the multipliers in the traditional systolic array and the bit width of the encoding-based multipliers in the proposed MAC design, respectively. The latter is determined by the search algorithm in Section \ref{sec:Methodology-MUL}. Although the bit width of the encoding-based multipliers is larger than that of the traditional one, the power consumption and the area cost of the MAC array exhibit significant advantage, because the logic to generate these intermediate bits is much simpler compared with the traditional multiplication. This advantage can be clearly seen from the last six columns in Table \ref{tab:generalHardwareResults} , where the column \textit{Trad.} and the column \textit{Prop.} show the results of power and area of the traditional MAC design and the proposed design, respectively.
The columns \textit{Red.} show the ratios of reduction
% \textcolor{blue}{
with respect to different size of MAC arrays.
Area and power of MAC arrays can be reduced by up to 48.79\% and 64.41\%, respectively.
% }.
% \textcolor{blue}{\sout{Besides, w}}
With the decreasing size of MAC array, the reduction of area and power consumption becomes smaller,
% \textcolor{blue}{
from 48.79\% and 64.41\% to 23.69\% and 43.10\%, respectively.
% }.
This phenomenon results from the fact that the bit-wise accumulators and decoders at the bottom of columns incur additional area and thus power consumption. When the MAC array has a small size, the incurred area and power cost contributes much to the total cost.

\begin{table}[t]
\setlength{\tabcolsep}{5pt}%2.5pt
\renewcommand{\arraystretch}{1.0}
\centering
\footnotesize
\caption{Inference accuracy of neural networks executed on the proposed MAC array}
\begin{tabular}{ccccc}
\toprule
\multirow{2}{*}{Model (Dataset)} &
  \multirow{2}{*}{32-FP} &
  \multicolumn{3}{c}{\begin{tabular}[c]{@{}c@{}}8-bit Uniform Quantization\end{tabular}} \\ 
  \cmidrule{3-5}
                             &         & Orig.    & Prop. & Acc. drop   \\ \midrule
        ResNet18 (Cifar-10 )  & 93.07\% &  93.01\% &  93.14\% & +0.13\%  \\ 
        ResNet20 (Cifar-100)  & 68.82\% &  68.40\% &  68.32\% & -0.08\%  \\ 
        ResNet50 (ImageNet)  & 76.15\% &  75.64\% &  75.46\% & -0.18\%  \\ 
     MobileNetV2 (Cifar-10)  & 93.91\% &  93.70\% &  93.74\% & +0.04\%  \\ 
     MobileNetV2 (Cifar-100) & 71.17\% &  70.72\% &  71.23\% & +0.51\%  \\ 
 EfficientNet-B0 (ImageNet)  & 77.69\% &  74.07\% &  69.96\% & -4.11\%  \\ 
  \bottomrule
\end{tabular}
\label{tab:generalAcc}
\end{table}

\subsubsection{Neural Networks on Encoding-based MAC Array}
%The proposed new MAC design can execute neural networks with high inference accuracy, as shown in 
Table \ref{tab:generalAcc} demonstrates the inference accuracy of neural networks executed on the proposed encoding-based MAC array. 
Neural networks together with the corresponding datasets are shown in the first column. 
The second column %of Table \ref{tab:generalAcc} is the 
shows the inference accuracy evaluated with 32-bit floating-point weights and input activations at software level.
The inference accuracy of neural networks with 8-bit uniform quantization 
% \textcolor{blue}{
and exact 8-bit multipliers
% } 
is shown in the third column.
The fourth column is their inference accuracy executed on the new MAC array.
% \textcolor{blue}{
% namely 93.14\%, 68.32\% and 75.46\% for three different models and datasets, respectively.
% }. 
%According to these columns, it is clear that 8-bit uniform quantization can nearly maintain the inference accuracy.
% with floating-point data. 
Due to the approximation in the proposed encoding technique, there is a slight accuracy loss, as shown in the fifth column. 
% \textcolor{blue}{
For ResNet20 (Cifar-100), ResNet50 (ImageNet), accuracy loss can be maintained within 0.2\%.
% 93.14\%, 68.32\% and 75.46\%
For ResNet18 (Cifar-10), MobileNetV2 (Cifar-10) and MobileNetV2 (Cifar-100), there is even a minor accuracy improvement after fine-tuning, 
% \textcolor{blue}{
i.e., from 93.01\% to 93.14\%, 93.70\% to 93.74\%, and 70.72\% to 71.23\%.
% }. 
While for large-scale EfficientNet-B0 (ImageNet), accuracy is recovered to 69.96\% after fine-tuning, leading to 4.11\% accuracy loss.
The main reason might be that the large number of layers in EfficientNet-B0 (81 \textit{Conv2d} layers) accumulates the approximate errors significantly. 
\subsubsection{Results of Simple Grid Search}
\begin{figure}[t]
% \vspace{8pt}
\centerline{\includegraphics[width=1\linewidth]{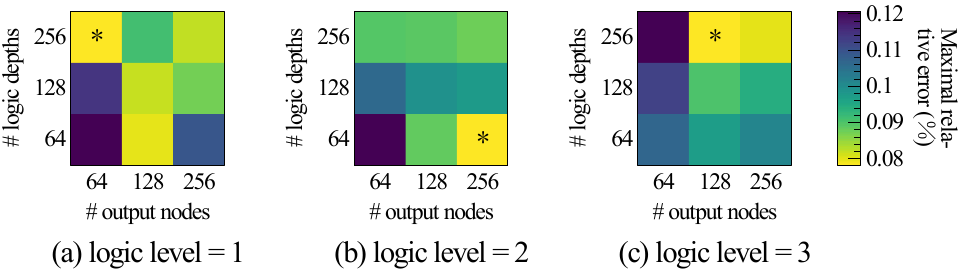}}
\caption{
% \textcolor{blue}{
The results of the simple grid search across different output nodes and logic depths for each logic level. The combinations of logic depths and output nodes marked with $\ast$ for each logic level are selected as the optimal hyperparameter settings in the subsequent CGP search sweep.
% }
}
\label{figCGP_search_config_sweep}
\end{figure}

As mentioned in Section \ref{sec:grid_search}, before the search of the CGP algorithm, 
the optimal hyperparameters such as the logic depth and the number of output nodes need to be determined for each logic level, respectively.  % to minimize the maximal relative error. %at different logic levels (1, 2, and 3), 
To achieve this goal, a simple grid search was applied to evaluate different combinations of such hyperparameters at different logic levels, as mentioned in Algorithm (\ref{Alg:grid_search}). 
% Each parameter has three value options, i.e., 64, 128, 256.  %conducted a simple grid search. T
%his search focused on two hyperparameters: the number of rows in hidden nodes and the number of output nodes, with possible values set at 64, 128, and 256. 
% For each combination, we can obtain an optimal encoding-based multiplier with a specified number of output bits, e.g., 64, that can achieve the smallest maximal relative error by search, as shown in \figname~\ref{figCGP_search_config_sweep}. 
As shown in \figname~\ref{figCGP_search_config_sweep}, the logic depth and the number of output nodes affects the maximal relative error of the multipliers obtained from search.
% \textcolor{blue}{\sout{significantly}}. 
%We consistently selected the best 64 nodes from the outputs nodes to serve as the 64-bit outputs of the encoding-based multiplier, using the CGP algorithm to minimize the maximal relative error between the encoding-based multiplier and an exact 8-bit multiplier. 
%The results of this simple grid search are illustrated in \figname~\ref{figCGP_search_config_sweep}.
For each logic level, the parameter combination  (marked with $\ast$ in \figname~\ref{figCGP_search_config_sweep})
achieving the smallest maximal relative error is selected 
% \textcolor{red}{\sout{for the phase 1 search }} 
for the subsequent CGP search sweep. 
% \textcolor{blue}{
% For example, when the number of logic levels is 2, we configure the number of logic depths and output nodes to 64 and 256, respectively.
% }
% \textcolor{blue}{
% As mentioned in Section \ref{sec:cgp}, the CGP search sweep is across all the possible combinations of output bit-width $M \in \{ 36$, $40$, $42$, $44$, $46$, $48$, $52$, $56$, $60$, $64\}$ and logic level $c\in\{1,2,3\}$ for each maximal relative error threshold $\varepsilon_{th}$ $\in$ $\{$$0.1\%$, $0.2\%$, $0.5\%$, $1\%$, $1.5\%$, $2\%$, $5\%$, $10\%$, $20\%$$\}$, respectively.
Namely, for each logic level $c\in\{1,2,3\}$, logic depth $r$ and output nodes $m$ are set to (256, 64), (64, 256) and (256, 128), respectively, according to \figname~\ref{figCGP_search_config_sweep}.

\begin{figure}[t]
% \vspace{8pt}
\centerline{\includegraphics[width=1\linewidth]{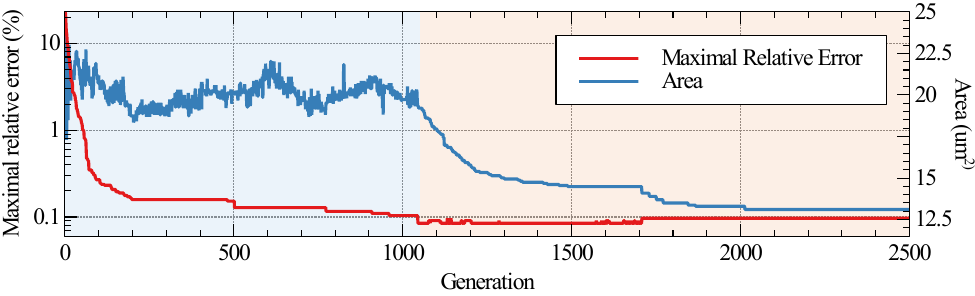}}
\caption{
% \textcolor{blue}{
An example of the CGP search process. From the 0th to 1045th generation, the CGP algorithm tries to reduce the maximal relative error to be less than a given threshold 0.1\%. From the 1046th to 2500th generation, the CGP algorithm targets to minimize the total area of the candidate multiplier.
% }
}
\label{figCGP_search_process}
\end{figure}

\subsubsection{Illustration of CGP Search Process}\label{sec:Illustration of CGP Search Process}
As mentioned in \ref{sec:cgp} and Algorithm (\ref{Alg:CGP_search}), 
the CGP search algorithm is adopted to search the encoding-based multiplier, 
using an open-source Python package \cite{schmidt20203889163}. 
% \textcolor{blue}{\sout{The search in this algorithm includes two phases, as}
An example of the search process in the CGP algorithm is
% }
shown in \figname~\ref{figCGP_search_process}.
%As depicted in \figname~\ref{figCGP_search_process}, the CGP (Cartesian Genetic Programming) algorithm operates in two phases. 
% \textcolor{blue}{\sout{In the first phase, the goal is}
From the 0th to 1045th generation in \figname~\ref{figCGP_search_process}, the first sub-function of the cost function in Equation (\ref{eq:cost_func}) works to guide the CGP algorithm
% }
to reduce the maximal relative error to be less than a given threshold, i.e. 0.1\%. Once this threshold is satisfied,
% \textcolor{blue}{\sout{the algorithm transitions to the second phase. The focus of the second phase is}
the second sub-function of the cost function works to guide the CGP algorithm
% }
to minimize the total area of the proposed encoding-based multiplier
% \textcolor{blue}{
from the 1046th to 2500th generation in \figname~\ref{figCGP_search_process}.
% }. 
%The total area is the sum of the areas of logic gates in all activated hidden nodes, with the area values derived from the 15nm NanGate technology library.
% The mutation rate is adjusted according to the maximal relative error of champions based on Table \ref{tab:CGP_mutate_rate}. 
%In the CGP algorithm, the mutation rate for each generation is dynamically adjusted based on the maximal relative error of the previous generation's champions.

\begin{figure}[t]
% \vspace{8pt}
\centerline{\includegraphics[width=1\linewidth]{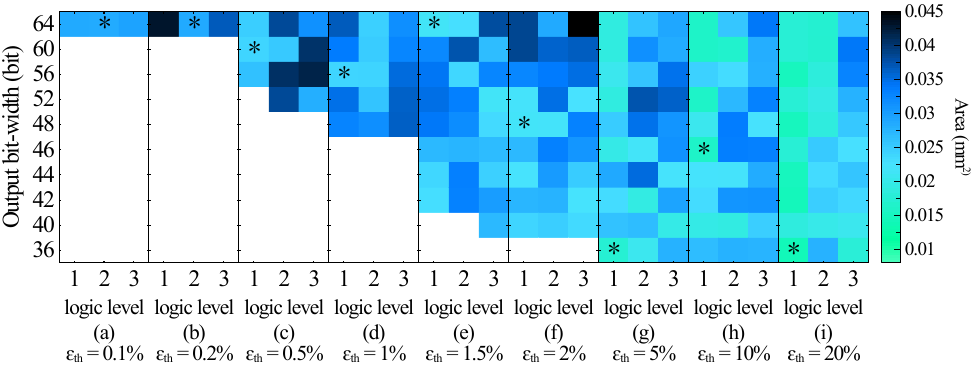}}
\caption{The results of the CGP search sweep across different logic levels and output bit-width for each maximal relative error threshold.
% \textcolor{blue}{
The combinations of output bit-width and logic levels of the multipliers that can achieve the smallest area under each given error threshold are marked with $\ast$.
% }
}
\label{fig_th_vs_output_bits}
\end{figure}

% \begin{table}[t]
% \setlength{\tabcolsep}{2pt}%2.5pt
% \renewcommand{\arraystretch}{1.2}
% \centering
% \footnotesize
% \caption{The best output bit-width  and logic level of searched multipliers under given maximal relative error thresholds.}
% \begin{tabular}{cccccccccc}
% \toprule
% \begin{tabular}[c]{@{}c@{}}Max. Relative\\ Error Threshold\end{tabular}
%    & 0.1\% & 0.2\% & 0.5\% & 1.0\% & 1.5\% & 2.0\% & 5.0\% & 10.0\% & 20.0\% \\ 
%    \midrule
% Output Bit-Wid. & 64    & 64    & 60    & 56    & 64    & 48    & 36    & 46     & 36     \\
% Logic Level     & 2     & 2     & 1     & 1     & 1     & 1     & 1     & 1      & 1      \\ 
% \bottomrule
% \end{tabular}
% \label{tab_th_BW_level}
% \end{table}

\subsubsection{Results of CGP Search Sweep}
\figname~\ref{fig_th_vs_output_bits} illustrates the results of the CGP search sweep at
% \textcolor{blue}{\sout{logic levels 1, 2, and 3}
different maximal relative error thresholds $\varepsilon_{th}$.
% }.
The white areas in the figure indicate that there are no encoding-based multipliers identified by the CGP algorithm that can meet the given threshold of maximal relative error, while the colored areas reflect the minimal area that the searched multipliers could achieve under the error threshold. 
Under a given error threshold, 
the output bit-width and logic levels
% \textcolor{blue}{significantly}
significantly
affect 
the area of the encoding-based multiplier circuits identified by the CGP algorithm. % varies with different output bit-width and logic levels, listed in 
% Table~\ref{tab_th_BW_level} listed the  % where the 
The
% \textcolor{blue}{combinations of}
combinations of
output bit-width and logic levels of the multipliers that can achieve the smallest area under each given error threshold 
% \textcolor{blue}{
are marked with $\ast$ in \figname~\ref{fig_th_vs_output_bits}.

\begin{figure}[t]
% \vspace{8pt}
\centerline{\includegraphics[width=1\linewidth]{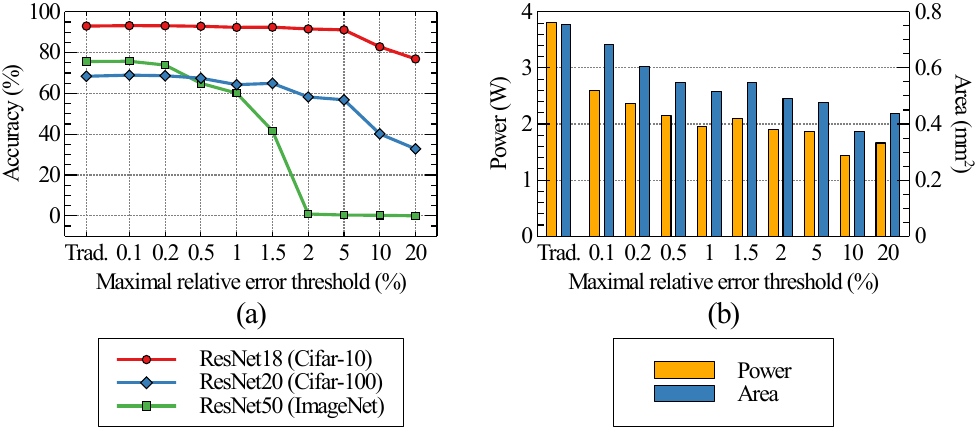}}
\caption{The relationship between the maximal relative error threshold and the inference accuracy of neural networks executed on the proposed MAC array (64$\times$64) and the power consumption as well as area. 
The MAC arrays are synthesized at 5 GHz clock frequency using Synopsys Design Compiler with 15nm NanGate Open-Cell Library \cite{OpenCellLibrary15nm}.
(a) Maximal relative error threshold vs. accuracy. (b) Maximal relative error threshold vs. power and area. %Accuracy (8-bit uniform quantization), power and area comparison between traditional (Trad.) and proposed general systolic array (size is 256 $\times$ 256).
}
\label{fig_retrain_accuracy_power_area}
\end{figure}

\subsubsection{Integration of Multipliers from CGP Search Sweep into Neural Networks}
The maximal relative error threshold reflects the
% \textcolor{blue}{\sout{fitness of}}
approximation between the encoding-based multiplier and the original multiplier. This approximation might degrade the inference accuracy of neural networks executing on such encoding-based MAC array. To select an encoding-based multiplier with a suitable maximal relative error threshold as our design, we evaluate the inference accuracy of neural networks, power consumption and area of the corresponding MAC array. The results are shown in \figname~\ref{fig_retrain_accuracy_power_area}. 
It is evident that as the maximal relative error threshold increases, accuracy, power consumption and area decreases. 
%Compared to the baseline, 
To keep a minimal accuracy loss for all the neural networks, we select 
the encoding-based multiplier with the maximal relative error threshold 0.1\% as our design
% \textcolor{blue}{
(marked with $\ast$ in \figname~\ref{fig_th_vs_output_bits}(a)).
% }.
%and called the \textit{critical threshold} later. Notably, the value of the critical threshold is inversely related to the difficulty of the task: the harder the task, the smaller the critical threshold. For instance, for the simple Cifar-10 classification task, the critical threshold is 1.5\%, whereas for the challenging Cifar-100 task, it is 0.5\%. 

\begin{figure}[t]
% \vspace{8pt}
\centerline{\includegraphics[width=1\linewidth]{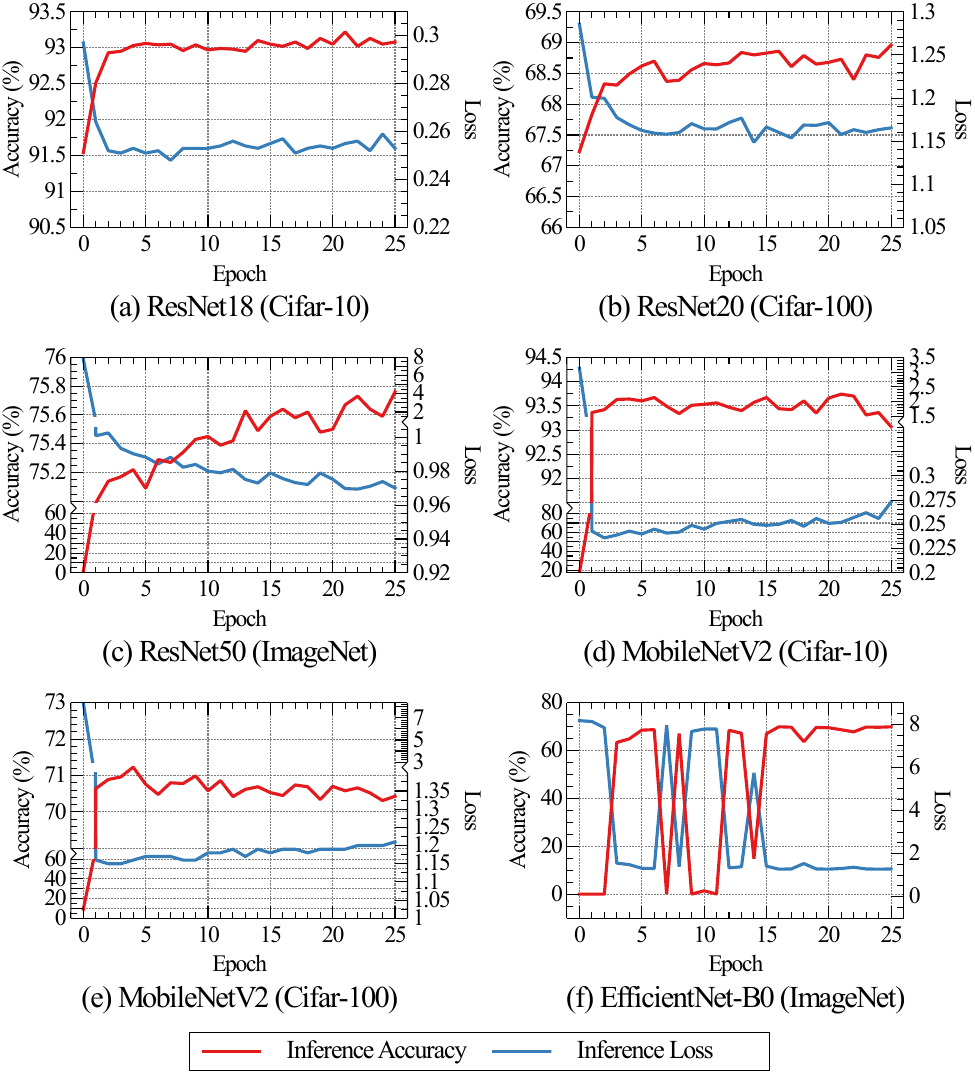}}
\caption{
The inference accuracy and loss function for each epoch of the fine-tuned model with 0.1\% maximal relative error threshold. 
% \textcolor{blue}{
The 0-th epoch indicates the accuracy without fine-tuning, i.e. 91.54\%, 67.23\%, 1.38\%, 17.17\%, 9.21\% and 0.10\% for (a), (b), (c), (d), (e) and (f), respectively.
% }
% (a) ResNet18 (Cifar-10). (b) ResNet20 (Cifar-100). (c) ResNet50 (ImageNet)
%Illustration of proposed binary search algorithm for the encoding of the product of 8-bit uniform quantization levels. Left: RMSE vs. iterations of search algorithm. Right: RMSE vs. the number of random samples.
}
\label{fig_retrain_process}
\end{figure}

% \begin{table}[t]
% \centering
% \footnotesize
% \caption{Inference accuracy before, during and after fine-tuning.}
% \begin{tabular}{cccc}
% \toprule
%   \multicolumn{1}{c}{\begin{tabular}[c]{@{}c@{}}Epochs\end{tabular}}
% & \multicolumn{1}{c}{\begin{tabular}[c]{@{}c@{}}MobileNetV2 \\(Cifar-10)\end{tabular}}
% & \multicolumn{1}{c}{\begin{tabular}[c]{@{}c@{}}MobileNetV2 \\(Cifar-100)\end{tabular}}
% & \multicolumn{1}{c}{\begin{tabular}[c]{@{}c@{}}EfficientNet-B0 \\(ImageNet)\end{tabular}}
% \\ \midrule
% w/o fine-tune & 17.17\% &  9.21\% &   0.10\%  \\\midrule
%       1       & 93.36\% & 70.64\% &   0.11\%  \\
%       2       & 93.36\% & 70.64\% &   0.11\%  \\
%       3       & 93.36\% & 70.64\% &  63.43\%  \\
%       5       & 93.60\% & 70.76\% &  67.17\%  \\
%       10      & 93.74\% & 71.23\% &  68.77\%  \\\midrule
%       % 15      & 93.67\% & 70.53\% &       \%  \\
%       % 20      & 93.66\% & 70.70\% &       \%  \\
%       % 25      & 93.74\% & 71.23\% &       \%  \\\midrule
%       32-FP   & 93.91\% & 71.17\% &  77.69\%  \\
% \bottomrule
% \end{tabular}
% \label{tab:finetune-mobilenet-efficientnet-acc-2}
% \end{table}

\subsubsection{Discussion of Neural Network Fine-tuning}
During the evaluation of neural networks, fine-tuning technique is used to 
boost the inference accuracy on the encoding-based MAC array. 
The inference accuracy and loss in each epoch during
% \textcolor{blue}{the}
the fine-tuning process are illustrated in \figname~\ref{fig_retrain_process}. 
The $0$-th epoch indicates the initial accuracy and loss before fine-tuning. 
It is observed that the initial accuracies for ResNet18 (Cifar-10) and ResNet20 (Cifar-100) are high, while that of ResNet50 (ImageNet), 
% \textcolor{blue}{
MobileNetV2 (Cifar-10), MobileNetV2 (Cifar-100) and EfficientNet-B0 (ImageNet), 
% }
is quite low, namely 1.38\%, 17.17\%, 9.21\% and 0.10\%, respectively. 
It is likely due to its higher task complexity and the larger number of convolution layers in ResNet-50, 
% \textcolor{blue}{
MobileNetV2 and EfficientNet-B0, 
% }
which accentuate the cumulative effect of approximation errors. 
Fortunately, the accuracies for all these tasks are quickly improved after fine-tuning. Particularly for ResNet50 (ImageNet) 
% \textcolor{blue}{
and EfficientNet-B0 (ImageNet), 
% }
as mentioned in Section \ref{sec:ex_setup}, we only used 1\% of the training data, while enabling the accuracy 
% \textcolor{blue}{
of ResNet50
% } 
to recover from 1.38\% to 75\% after just one epoch, 
% \textcolor{blue}{
and of EfficientNet-B0 from 0.10\% to 63.43\% after two fine-tuning epochs. 
Another observation is that the inference accuracy of EfficientNet-B0 (ImageNet) 
fluctuated drastically between 0.1\% and 68\% in the early epochs of fine-tuning, and then stabilized gradually around 69\% after about 15 epochs.
EfficientNet-B0 is deeper than others (81 \textit{Conv2d} layers), so the approximate errors are easy to be accumulated. 
Thus, using only 1\% of the training data speeds up fine-tuning but might be insufficient to quickly stabilize the generalization performance of the large-scale EfficientNet-B0. 

\subsubsection{Minor Change to Position Weights}
\begin{figure}[t]
% \vspace{8pt}
\centerline{\includegraphics[width=1\linewidth]{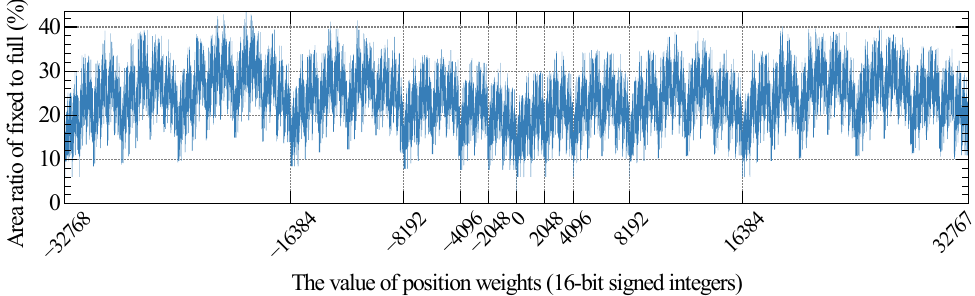}}
\caption{The area ratio of fixed multiplier with different position weight values to full multiplier. The area 
% \textcolor{blue}{is}
is
obtained by synthesizing the circuits in 15nm NanGate cell library. %The signed 15-bit position weights are used to simplify full multipliers.
}
\label{fig_area_of_fixed_multipliers}
\end{figure}

\begin{figure}[t]
\centerline{\includegraphics[width=1\linewidth]{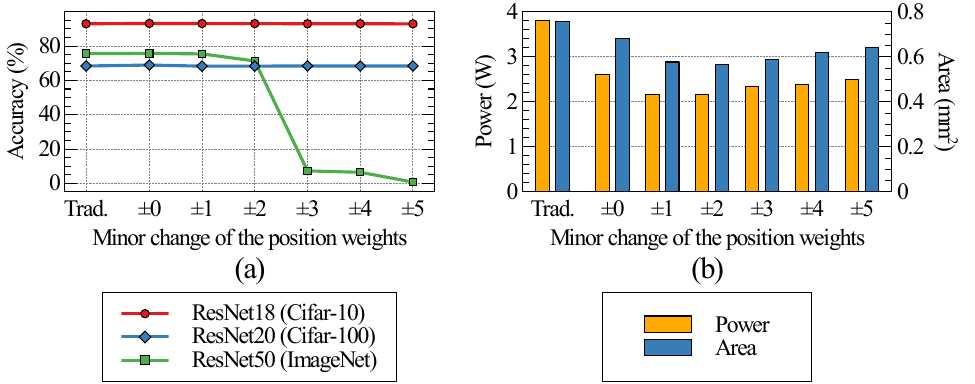}}
\caption{
The inference accuracy, power consumption and area vs. the minor change of position weights on the proposed MAC array (64$\times$64). 
The MAC arrays are synthesized at 5 GHz clock frequency using Synopsys Design Compiler with 15nm NanGate Open-Cell Library \cite{OpenCellLibrary15nm}.
 (a) The minor change of position weights affect inference accuracy of neural networks. (b) 
 The minor change of position weights bring reduction in area and power. 
 %The value of minor change of position weights vs. inference accuracy, power and area.
}
\label{fig_change_position_weights}
\end{figure}

%To implement the encoding-based MAC array, %full multipliers are required to realize the multiplication of the output bit of the encoding-based multiplier and the position weight. 
%The position weights are obtained by approximating the encoding-based multiplier as accuracy as possible, so that they are the same for different neural networks. 
% \textcolor{blue}{
As mentioned above, we select the encoding-based multiplier with the configuration that $c=2$, $M=64$, $\varepsilon_{th}=0.1\%$ (marked with $\ast$ in \figname~\ref{fig_th_vs_output_bits}(a)) as our design.
The position weights of this design are initially determined by the CGP algorithm with Equation (\ref{eq:lst_PositionWeights_v3}).
The position weights are identical for different neural networks, so they can be fixed to simplify the logic of multipliers and adder tree after ACC units in \figname~\ref{fig:ourMAC}.
% }
However, different position weight values lead to different area of simplified multipliers, as shown in \figname~\ref{fig_area_of_fixed_multipliers}.
Inspired by this difference, we adjusted the values of position weights within a specified range, attempting to look for alternatives that result in a smaller area of simplified multiplier to replace the original ones.
% \textcolor{blue}{
For example, the original value of a position weight determined by CGP algorithm is 254, which can be used to simplify the multiplier. If we apply a minor change to the value, i.e., changing the value from 254 to 256, we can obtain a further simplified multiplier with a smaller area.
% }
The adjusted position weights were used to evaluate the inference accuracy, power consumption and area.
\figname~\ref{fig_change_position_weights} presents the evaluation results. As the values of position weights are adjusted, accuracy remains relatively stable for the Cifar-10 and Cifar-100 classification tasks, but significantly degrades for the more challenging ImageNet task. 
By limiting the adjustments to position weights within $\pm$1, the accuracy across all three tasks remains stable, and there is a significant reduction in the area and power consumption. Therefore, this circuit configuration has been adopted as our final design, with detailed data on accuracy, area, and power consumption provided in Table~\ref{tab:generalHardwareResults} and Table~\ref{tab:generalAcc}.

% \subsection{Comparison Study}
\subsection{Comparison with Related Approaches}
\subsubsection{
% Hardware and Accuracy Comparison with 
Approximate Systolic Arrays}\label{Accuracy and Hardware Comparison with Approximate Systolic Arrays}
\begin{figure}[t]
\centerline{\includegraphics[width=1\linewidth]{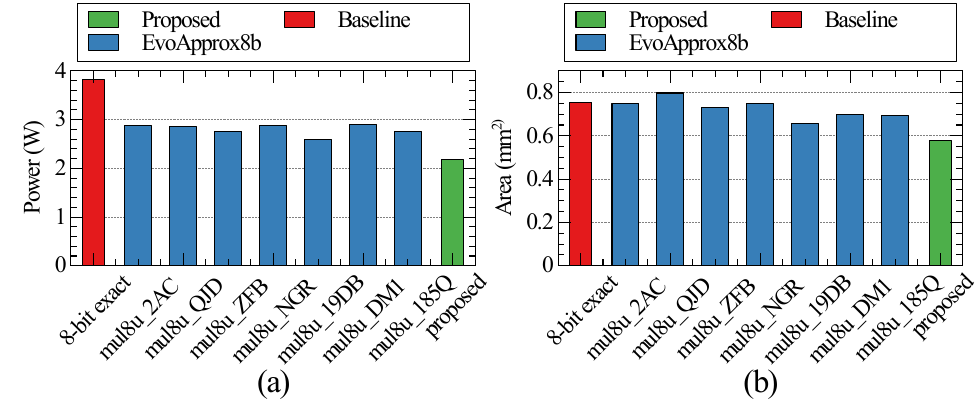}}
\caption{
Comparison of (a) power consumption and (b) area for MAC arrays (64$\times$64) between the proposed MAC array and the MAC array constructed by the approximate multipliers in EvoApprox8b open source Library \cite{mrazek2017evoapprox8b, onlineEvoApprox8b}. These approximate multipliers are used and applied to neural networks in \cite{pinos2023acceleration} and \cite{mrazek2020libraries}. The MAC arrays are synthesized at 
% \textcolor{blue}{\sout{1 GHz}}
5 GHz
clock frequency using Synopsys Design Compiler with 15nm NanGate Open-Cell Library \cite{OpenCellLibrary15nm}.
}
\label{fig_compare_area_power}
\end{figure}

To compare with other approximate computing techniques, we selected 7 different approximate multipliers from the open-source EvoApprox8b library \cite{mrazek2017evoapprox8b} to construct 64$\times$64 MAC arrays (the adders are still accurate), and compared them with the MAC array based on the encoding-based multiplier proposed in this paper in terms of power consumption and area. These 7 approximate multipliers (including mul8u\_2AC, mul8u\_QJD, mul8u\_ZFB, mul8u\_NGR, mul8u\_19DB, mul8u\_DM1, and mul8u\_185Q) have been applied in neural networks and achieved acceptable accuracy as reported in \cite{pinos2023acceleration} and \cite{mrazek2020libraries}. We built the MAC arrays using the online available Verilog codes of these multipliers \cite{onlineEvoApprox8b}, synthesized them using the 15 nm NanGate Open-Cell Library \cite{OpenCellLibrary15nm} and Synopsys Design Compiler at 
% \textcolor{blue}{\sout{1 GHz}}
5 GHz clock frequency, and obtained power consumption and area for comparison. 
% \textcolor{red}{First should write about what is good.}
The comparison results are presented in \figname~\ref{fig_compare_area_power}. 
The proposed design has area and power advantages over the approximate multipliers from the EvoApprox8b library when they are integrated into MAC arrays. This is mainly because in traditional MAC arrays, the registers that store intermediate partial sums still dominate the area and power consumption. However, the MAC array built with the encoding-based multiplier proposed in this paper, due to its fewer logic levels and smaller delay, enables highly parallelized computation using bit-wise accumulation, eliminating the intermediate registers used to store partial sums, thereby cutting down on area and power consumption. 

\begin{table}[t]
\setlength{\tabcolsep}{1.3pt}%2.5pt
\renewcommand{\arraystretch}{1.0}
\centering
% \tiny
\fontsize{7.5}{7} \selectfont
\caption{Comparison between the proposed design and systolic arrays based on approximate and accurate multipliers and accumulators.}
\begin{tabular}{cccccccc}
\toprule
   \multicolumn{1}{c}{\begin{tabular}[c]{@{}c@{}}Multiplier\end{tabular}}
& \multicolumn{1}{c}{\begin{tabular}[c]{@{}c@{}}Adder\end{tabular}}
% & \multicolumn{1}{c}{\begin{tabular}[c]{@{}c@{}}Config\end{tabular}}
& \multicolumn{1}{c}{\begin{tabular}[c]{@{}c@{}}Area (mm$^2$)\\64$\times$64\end{tabular}}
& \multicolumn{1}{c}{\begin{tabular}[c]{@{}c@{}}Power (W)\\64$\times$64\end{tabular}}
& \multicolumn{1}{c}{\begin{tabular}[c]{@{}c@{}}ResNet18 \\(Cifar-10)\end{tabular}}
& \multicolumn{1}{c}{\begin{tabular}[c]{@{}c@{}}ResNet20 \\(Cifar-100)\end{tabular}}
& \multicolumn{1}{c}{\begin{tabular}[c]{@{}c@{}}ResNet50 \\(ImageNet)\end{tabular}}
\\ \midrule
exact  & exact  & 0.76 & 3.82 & 93.01\% & 68.40\% &  75.64\% \\
 exact  & approx & 0.58 & 2.61 & 91.77\% & 65.55\% &  71.54\% \\
 approx & exact  & 0.68 & 2.59 & 92.74\% & 67.53\% &  74.94\% \\
approx & approx & 0.54 & 2.02 & 90.34\% & 63.23\% &  68.37\% \\
 ours   & exact  & 0.58 & 2.17 & 93.14\% & 68.32\% &  75.46\% \\
\bottomrule
\end{tabular}
\label{tab:systolic-approx-approx-2}
\end{table}

% \textcolor{blue}{
We also conducted the  comparison between the proposed MAC array and systolic arrays using approximate multipliers and approximate adders. We selected ``mul8s\_1L2H'' from \cite{mrazek2017evoapprox8b} as the approximate multiplier, and the approximate adder in \cite{ACA-CSU} for this comparison. Approximate adders from \cite{mrazek2017evoapprox8b} are not used, because the maximum input bit-width in \cite{mrazek2017evoapprox8b} is only 16, which is not sufficient to maintain the accuracy. We used the selected approximate multiplier and adder to build a 64$\times$64 systolic array and evaluated the area and power consumption. We compiled the C codes of the approximate multiplier and adder, and then embed them to Pytorch to fine-tune the neural networks and test the accuracy. The results are shown in Table \ref{tab:systolic-approx-approx-2}. 
%Let [E/A] represent the configuration of the exact multiplier and approximate adder in systolic array. Compared with [E/E], [E/A] and [A/E] configurations, 
According to this comparison, the proposed design achieves better area, power and accuracy generally. Compared with the systolic array with approximate multipliers and adders, the area and power of the proposed design are slightly higher, but the accuracy is much better.
% }

\begin{table}[t]
\setlength{\tabcolsep}{3.5pt}%2.5pt
\renewcommand{\arraystretch}{1.0}
\centering
% \tiny
\fontsize{7.5}{7} \selectfont
\caption{Comparison of exact, approximate, and proposed multipliers combined with the proposed bit-wise accumulation. }
\begin{tabular}{ccccccc}
\toprule
  \multicolumn{1}{c}{\begin{tabular}[c]{@{}c@{}}Architecture\end{tabular}}
& \multicolumn{1}{c}{\begin{tabular}[c]{@{}c@{}}Multiplier\end{tabular}}
& \multicolumn{1}{c}{\begin{tabular}[c]{@{}c@{}}Adder\end{tabular}}
& \multicolumn{1}{c}{\begin{tabular}[c]{@{}c@{}}Area (mm$^2$)\\64$\times$64\end{tabular}}
% & \multicolumn{1}{c}{\begin{tabular}[c]{@{}c@{}}Power (W)\\64$\times$64\end{tabular}}
& \multicolumn{1}{c}{\begin{tabular}[c]{@{}c@{}}Timing\end{tabular}}
& \multicolumn{1}{c}{\begin{tabular}[c]{@{}c@{}}Slack (ps)\end{tabular}}
\\ \midrule
proposed & ours   & exact & 0.58 & MET      & 5.96   \\
proposed & exact  & exact & 0.60 & VIOLATED & -25.59 \\
proposed & approx & exact & 0.59 & VIOLATED & -5.64  \\
\bottomrule
\end{tabular}
\label{tab:prop-approx-approxtext-2}
\end{table}

\subsubsection{
% Hardware Comparison with 
Bit-Wise Accumulation with Exact or Approximate Multipliers}\label{Hardware Comparison with Bit-Wise Accumulation of Exact and Approximate Multipliers}
% \textcolor{blue}{
We conducted another comparison to combine different multipliers with the bit-wise accumulation in the proposed approach. 
For this comparison, we compared exact, approximate (``mul8s\_1L2H'' from \cite{mrazek2017evoapprox8b}), and proposed multipliers combined with the proposed bit-wise accumulation adders. 
The results are shown in Table~\ref{tab:prop-approx-approxtext-2}. According to this comparison, these designs have similar area overhead,
but the version with exact and approximate multiplers did not meet the timing requirement as shown
in the last two rows in Table~\ref{tab:prop-approx-approxtext-2}, because these multipliers are much more complicated than the proposed
multipliers.
% }

\begin{table}[t]
\setlength{\tabcolsep}{2.2pt}%2.5pt
\renewcommand{\arraystretch}{1.0}
\centering
% \tiny
\fontsize{7.5}{7} \selectfont
\caption{Power and area comparison between MAC units based on 2's complement, signed magnitude  and the proposed enconding.}
\begin{tabular}{cccccccc}
\toprule
 \multicolumn{2}{c}{\begin{tabular}[c]{@{}c@{}}Multiplier\end{tabular}}  &
& \multicolumn{2}{c}{\begin{tabular}[c]{@{}c@{}}Accumulator\end{tabular}} &
& \multirow{2}{*}{\begin{tabular}[c]{@{}c@{}}Area (mm$^2$)\\64$\times$64\end{tabular}}
& \multirow{2}{*}{\begin{tabular}[c]{@{}c@{}}Power (W)\\64$\times$64\end{tabular}}
\\ \cmidrule{1-2}\cmidrule{4-5}
           input       & output       & & input        & output        & &      &      \\ \midrule
  2's comp.   & 2's comp.    & & 2's comp.    & 2's comp.     & & 0.76 & 3.82 \\
  signed mag. & signed mag.  & & signed mag.  & signed mag.   & & 0.88 & 3.22 \\
  2's comp.   & ours         & & ours         & 2's comp.     & & 0.58 & 2.17 \\
\bottomrule
\end{tabular}
\label{tab:power-area-sig-mag-MAC-2}
\end{table}

\subsubsection{
% Hardware Comparison with 
Signed Magnitude Systolic Arrays}\label{Hardware Comparison with Signed Magnitude Systolic Arrays}
% \textcolor{blue}{
We also compared the area and power consumption of systolic arrays where MAC units were implemented based on 2's complement, signed magnitude and the propsoed encoding. 
% we implemented MAC units based on signed magnitude format.
% according to the  online simulation link ``\url{http://hdlbits.01xz.net/wiki/Iverilog?load=px3nzv}''. 
The comparison results are shown in 
%Area and power of 64$\times$64 MAC arrays based on 2's complement, signed magnitude and our encoding formats are compared in 
Table \ref{tab:power-area-sig-mag-MAC-2}. 
According to this table, 
the systolic array based on signed magnitude format achieves lower power, but larger area than the one with 2's complement. The proposed encoding can achieve the smallest 
area overhead and power consumption.
% }

\subsubsection{
% Hardware and Accuracy Comparison with 
Low Bit-Width Integer Quantization}\label{Hardware and Accuracy Comparison with Integer Quantization}
% \textcolor{blue}{
We also implemented a 64$\times$64 MAC array with the corresponding quantized MAC units and evaluated the resulting area and power consumption to compare with the propose technique with heavily quantized networks. 
The results are shown in Table~\ref{tab:inttext}, where the last row indicates the resutls of the proposed method. 
To extensively evaluate the quantization configurations, we swept all combinations of INT2/INT4/INT6/INT8 activations and weights, and fine-tuning technique was also applied to enhance the inference accuracy. 
%Assume [INT$n$/INT$n$] represents the quantization scheme for $n$-bit activations and $n$-bit weights. 
For ResNet 18 (Cifar-10), accuracy can be well maintained on INT6-INT4, INT6-INT6, INT6-INT8, INT8-INT4, INT8-INT6, and INT8-INT8. Area and power consumption of INT6-INT4 and INT8-INT4 are indeed smaller than our proposed method. But for ResNet 20 (Cifar-100), accuracy can be only well maintained on INT8-INT6 and INT8-INT8, and area and power consumption are both larger than ours. For ResNet 50 (ImageNet), accuracy can be only maintained on INT8-INT8 and the implementation of the corresponding systolic arrays incurs the largest power and area. Generally, the proposed method can balance accuracy, area overhead and power consumption for deep neural networks. 

\begin{table}
\setlength{\tabcolsep}{2.1pt}%2.5pt
\renewcommand{\arraystretch}{1.0}
\centering
% \tiny
\fontsize{7.5}{7} \selectfont
\caption{Accuracy comparison of neural networks on various quantization bit-widths for weights and activations after fine-tuning, as well as the corresponding power and area with the quantization configuration. The last row indicates the results with the proposed method, and the others indicate the results with the systolic array.}
\begin{tabular}{cccccc}
\toprule
   \multicolumn{1}{c}{\begin{tabular}[c]{@{}c@{}}Activation-Weight\end{tabular}}  
% & \multicolumn{1}{c}{\begin{tabular}[c]{@{}c@{}}Weights\end{tabular}} 
& \multicolumn{1}{c}{\begin{tabular}[c]{@{}c@{}}ResNet18 \\(Cifar-10)\\32FP: 93.07\%\end{tabular}}
& \multicolumn{1}{c}{\begin{tabular}[c]{@{}c@{}}ResNet20 \\(Cifar-100)\\32FP: 68.82\%\end{tabular}}
& \multicolumn{1}{c}{\begin{tabular}[c]{@{}c@{}}ResNet50 \\(ImageNet)\\32FP: 76.15\%\end{tabular}}
& \multicolumn{1}{c}{\begin{tabular}[c]{@{}c@{}}Area \\(mm$^2$)\\64$\times$64\end{tabular}}
& \multicolumn{1}{c}{\begin{tabular}[c]{@{}c@{}}Power \\(W)\\64$\times$64\end{tabular}}

\\ \midrule
 INT2-INT2 & 10.00\% &  1.00\%  &  0.10\% & 0.17  & 0.80  \\
 INT2-INT4 & 65.65\% &  6.21\%  &  0.12\% & 0.23  & 1.06  \\
 INT2-INT6 & 66.34\% & 12.82\%  &  0.15\% & 0.26  & 1.24  \\
 INT2-INT8 & 66.77\% & 14.95\%  &  0.13\% & 0.30  & 1.46  \\ \midrule
 INT4-INT2 & 86.83\% &  1.00\%  &  0.10\% & 0.23  & 1.06  \\
 INT4-INT4 & 90.88\% & 54.70\%  & 44.29\% & 0.31  & 1.31  \\
 INT4-INT6 & 90.70\% & 57.32\%  & 51.66\% & 0.37  & 1.82  \\
 INT4-INT8 & 91.30\% & 60.13\%  & 52.06\% & 0.49  & 2.59  \\ \midrule
 INT6-INT2 & 90.66\% &  5.10\%  &  0.10\% & 0.26  & 1.17  \\
 INT6-INT4 & 92.64\% & 58.99\%  & 46.97\% & 0.37  & 1.59  \\
 INT6-INT6 & 93.08\% & 65.52\%  & 66.70\% & 0.50  & 2.55  \\
 INT6-INT8 & 92.91\% & 65.85\%  & 71.75\% & 0.61  & 2.97  \\ \midrule
 INT8-INT2 & 90.36\% &  1.00\%  &  0.10\% & 0.30  & 1.38  \\
 INT8-INT4 & 92.86\% & 60.95\%  & 47.20\% & 0.45  & 1.91  \\
 INT8-INT6 & 92.97\% & 67.95\%  & 71.92\% & 0.60  & 2.46  \\
 INT8-INT8 & 93.13\% & 68.40\%  & 75.64\% & 0.76  & 3.80  \\ \midrule
 INT8-INT8 & 93.14\% & 68.32\%  & 75.46\% & 0.58  & 2.17  \\ % \midrule
% Systolic & FP32 & FP32 & 93.07\% & 68.82\%  & 76.15\% & -       & -      \\
\bottomrule
\end{tabular}
\label{tab:inttext}
\end{table}

\section{Conclusion}
\label{sec:conclusion}
In this
paper, we have proposed a novel digital MAC design based on encoding. 
%In this new design, the multipliers are replaced by simple logic gates to project the results to a wide bit representation. These bits carry individual position weights, which can be trained for specific neural networks to enhance inference accuracy. The wide bits and the corresponding position weights are then used to calculate the outputs of neurons by bit-wise weighted accumulation in an MAC
%array. %These outputs at neurons are in the original formats specified by the neural networks with either uniform or non-uniform quantization, so that the proposed design is compatible with existing computing platforms. 
%Since the multiplication function is
%replaced by simple logic projection, the critical paths in the resulting circuits become much shorter. Correspondingly, pipelining stages in the MAC array can be reduced significantly, which can enhance area efficiency and computational performance simultaneously. 
With this technique, the complex logic in traditional multipliers can be replaced with 
simpler logic to reduce the critical path and area significantly. 
The position weights allow a bit-wise accumulation to calculate the addition result. Correspondingly, pipelining stages between rows of multipliers can be reduced to lower area cost further. With this new design, area and power consumption of MAC array can be reduced by up to 48.79\% and 64.41\%, respectively, compared with traditional design while the inference accuracy is still maintained. 
%\textcolor{blue}{\sout{Future work will study the tradeoff between the bit width at the output of the multipliers and the complexity of the single-level 
%as well as the multi-level logic implementation in the simplified multipliers. } Future work will study how to represent the outputs of multiplication using wide bit encodings with position weights without approximation error, and explore the trade-offs between area, delay and power consumption of the accurate encoding-based multipliers and MAC arrays.

%while the accuracy of the neural networks can still be well maintained.
%The proposed design has been synthesized and verified
%by ResNet18-Cifar10, ResNet20-Cifar100 and ResNet50-ImageNet.
%The experimental results confirmed the reduction of circuit area by
%up to 70.18% and the reduction of power consumption of executing
%DNNs by up to 79.63%, while the accuracy of the neural networks
%can still be well maintained
%%

\let\oldthebibliography=\thebibliography
\let\endoldthebibliography=\endthebibliography
\renewenvironment{thebibliography}[1]{%
\begin{oldthebibliography}{#1}%
\setlength{\itemsep}{0.15ex}%
\fontsize{7.2pt}{1}\selectfont
%\vskip 1pt
%\scriptsize
%\small
%\footnotesize
%\reffontsize
\newlength{\mylength}
\setlength{\mylength}{7.2pt}
\setlength{\baselineskip}{\baselinestretch\mylength}
}%
{%
\end{oldthebibliography}%
}

\bibliographystyle{IEEEtran}
%\bibliographystyle{ACM-Reference-Format}
%\bibliography{IEEEabrv,CONFabrv,bibfile}
\normalem
\bibliography{sample-base}
%\newpage

% your own custom macro containing the $\backslash${\tt{includegraphics}} command to make things
 

%\bf{If you include a photo:}\vspace{-33pt}
%\begin{IEEEbiography}%%[{\includegraphics[width=1in,height=1.25in,clip,keepaspectratio]{fig1}}]
%{Michael Shell}
%Use $\backslash${\tt{begin\{IEEEbiography\}}} and then for the 1st argument use $\backslash${\tt{includegraphics}} to declare and link the author photo.
%Use the author name as the 3rd argument followed by the biography text.
%\end{IEEEbiography}


%Use $\backslash${\tt{begin\{IEEEbiographynophoto\}}} and the author %name as the argument followed by the biography text.

\begin{IEEEbiography}[{\includegraphics[width=1in,height=1.25in,clip,keepaspectratio]{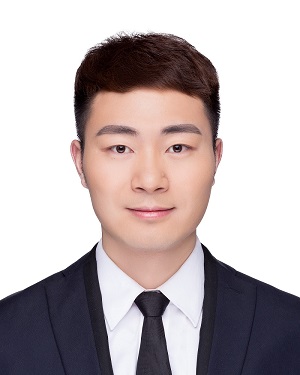}}]{Bo Liu} received the B.S. degree in microelectronics from Jilin University (JLU), Changchun, China, in 2016, 
and the M.Eng. degree in microelectronics and solid-state electronics from Southeast University (SEU), Nanjing, China, in 2020. 
He is currently pursuing the Dr.-Ing. degree with the Chair of Electronic Design Automation, Technical University of Munich, Munich, Germany. 
His research interests hardware accelerators for AI algorithms and systems, AI computing with emerging systems, and AI applications in optical communication systems. 
\end{IEEEbiography}

\begin{IEEEbiography}[{\includegraphics[width=1in,height=1.25in,clip,keepaspectratio]{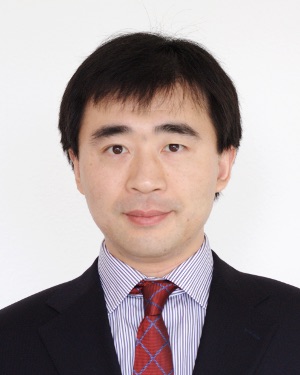}}]{Bing Li} (Senior Member, IEEE) received the Dr.-Ing. degree from the Technical University of Munich, Munich, Germany, in 2010. He is currently a professor with the Digital Integrated Systems Group, University of Siegen, Germany. His research interests include high-performance and low-power design, emerging computing systems, 
and machine learning for electronic design automation. 
He has served on the Technical Program Committee of several conferences, including DAC, ICCAD, DATE, ASP-DAC, and MLCAD.
\end{IEEEbiography}

\begin{IEEEbiography}[{\includegraphics[width=1in,height=1.25in,clip,keepaspectratio]{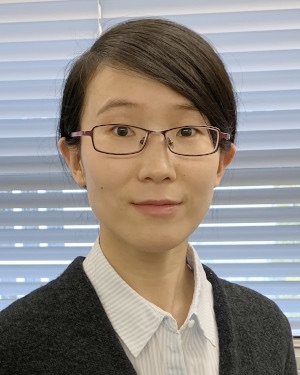}}] {Grace Li Zhang} (Member, IEEE) received the Dr.-Ing. degree from the Technical University of Munich, Munich, Germany in 2018.
She is currently an Assistant Professor with the Hardware for Artificial Intelligence Group, Technical University of Darmstadt, Darmstadt, Germany. Her research interests include hardware accelerators for AI algorithms and systems, AI computing with emerging devices, design methodologies for AI, explainability of AI, and neuromorphic computing. Dr. Zhang has served/is serving on the Technical Committee of several conferences, including DAC, ICCAD, ASP-DAC, and GLSVLSI.
\end{IEEEbiography}

\begin{IEEEbiography}[{\includegraphics[width=1in,height=1.25in,clip,keepaspectratio]{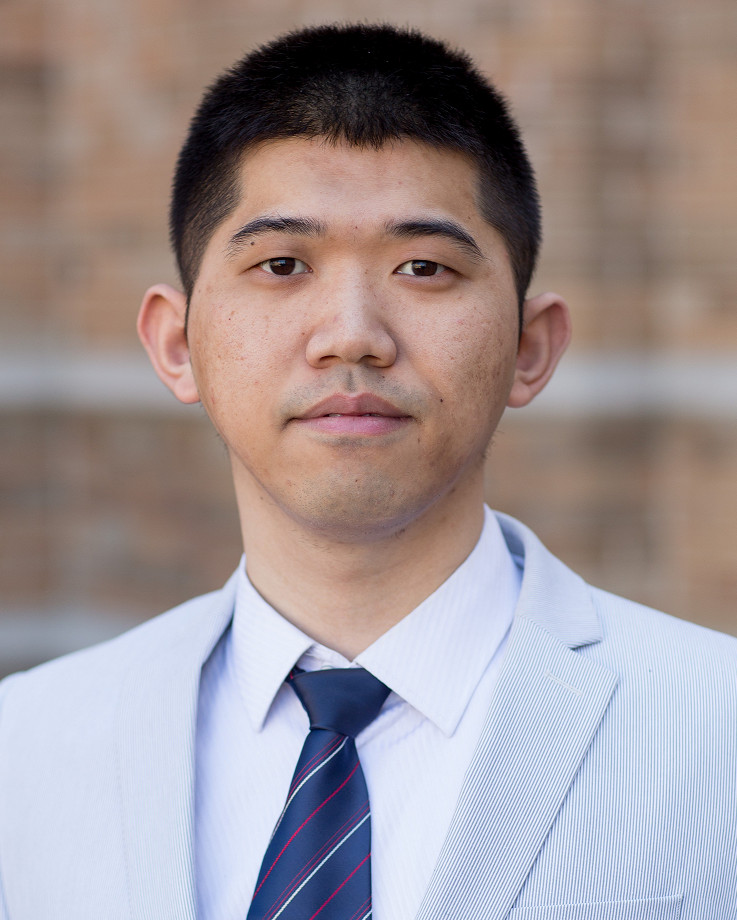}}] {Xunzhao Yin} (Member, IEEE) received the B.S. degree in electronic engineering from Tsinghua University, Beijing, China, in 2013, and the Ph.D. degree in computer science and engineering from the University of Notre Dame, Notre Dame, IN, USA, in 2019.

He is an Assistant Professor with the College of Information Science and Electronic Engineering, Zhejiang University, Hangzhou, China, and also with Zhejiang Lab, Hangzhou. He has published top journals and conference papers, including Nature
Electronics, IEEE TRANSACTIONS ON COMPUTERS, IEEE TRANSACTIONS ON COMPUTER-AIDED DESIGN OF INTEGRATED CIRCUITS AND SYSTEMS, IEEE TRANSACTIONS ON CIRCUITS AND SYSTEMS—I: REGULAR PAPERS, IEEE TRANSACTIONS ON ELECTRON DEVICES, DAC, IEDM, and Symposium on VLSI. His research interests include emerging circuit/architecture designs and novel computing paradigms with both CMOS and emerging technologies.

Dr. Yin has received the Best Paper Award Nomination of ICCAD 2020 and VLSI Test Symposium 2021, and the Zhejiang University 2020 Top 10 Academic Research Advancements Nomination.
\end{IEEEbiography}

\begin{IEEEbiography}[{\includegraphics[width=1in,height=1.25in,clip,keepaspectratio]{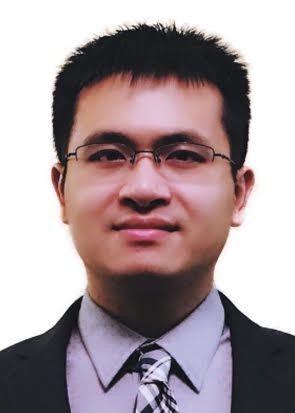}}] {Cheng Zhuo} 
% (Senior Member, IEEE) received the B.S. (Hons.) and M.S. degrees in electronic engineering from Zhejiang University, Hangzhou, China, in 2005 and 2007, respectively, and the Ph.D. degree in computer science and engineering from the University of Michigan, Ann Arbor, MI, USA, in 2010.

% He is currently a Professor with the College of Information Science and Electronic Engineering, Zhejiang University. His current research interests include computing in memory, deep learning, hardware accelerator, and general VLSI EDA areas.

% Prof. Zhuo was the recipient of four best paper nominations, the
% ACM/SIGDA Technical Leadership Award and Meritorious Service Award, the JSPS Invitation Fellowship, and the Humboldt Fellowship for Experienced Researchers. He has served on the technical program and organization committees of many international conferences and as an Associate Editor for IEEE TRANSACTIONS ON COMPUTER-AIDED DESIGN OF INTEGRATED CIRCUITS AND SYSTEMS, ACM Transactions on Design Automation of Electronic Systems, and Integration (Elsevier).

(Senior Member, IEEE) received his B.S. and M.S. from Zhejiang University, Hangzhou, China, in 2005 and 2007. He received his Ph.D. from the University of Michigan, Ann Arbor, in 2010. 

Dr. Zhuo is currently Qiushi Distinguished Professor at Zhejiang University, where his research focuses on hardware intelligence, machine learning-assisted EDA, and low power designs. He has authored/coauthored more than 200 technical papers, receiving 5 Best Paper Awards and 6 Best Paper Nominations. He is the recipient of ACM/SIGDA Meritorious Service Award and Technical Leadership Award, JSPS Faculty Invitation Fellowship, Humboldt Research Fellowship, etc. 

Dr. Zhuo has served on the organization/technical program committees of many international conferences, as the area editor for Journal of CAD\&CG, and as Associate Editor for IEEE TCAD, ACM TODAES, and Elsevier Integration. He is IEEE CEDA Distinguished Lecturer, a senior member of IEEE, and a Fellow of IET.

\end{IEEEbiography}

\begin{IEEEbiography}[{\includegraphics[width=1in,height=1.25in,clip,keepaspectratio]{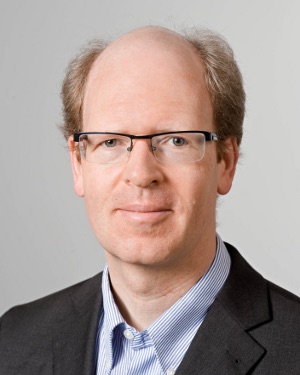}}] {Ulf Schlichtmann} (Senior Member, IEEE) received the Dipl.-Ing. and Dr.-Ing. degrees in electrical engineering and information technology from the Technical University of Munich (TUM), Munich, Germany, in 1990 and 1995, respectively.

He was with Siemens AG, Munich, and Infineon Technologies AG, Munich, from 1994 to 2003, where he held various technical and management positions in design automation, design libraries, IP reuse, and product development. He has been a Professor and the Head of the Chair of Electronic
Design Automation, TUM, since 2003, where he also served as the Dean of the Department of Electrical and Computer Engineering from 2008 to 2011, 
% and as an Associate Dean of Studies of International Studies from 2013 to 2022. 
and as Associate Dean for International Studies from 2013 to 2022.
His current research interests include computer-aided design of electronic circuits and systems, with an emphasis on designing reliable and robust systems. In recent years, he has increasingly worked on emerging technologies, such as microfluidic biochips and optical interconnect.

Prof. Schlichtmann is an Elected Member of TUM’s Academic Senate and Board of Trustees since 2016. He is a member of the German National Academy of Science and Engineering 
% and serves on many advisory boards.
and serves on various advisory boards.
\end{IEEEbiography}

\vfill

\end{document}